\documentstyle[11pt]{article}
\newcounter{subequation}[equation]

\makeatletter

\def\thesubequation{\theequation\@alph\c@subequation}
\def\@subeqnnum{{\rm (\thesubequation)}}
\def\slabel#1{\@bsphack\if@filesw {\let\thepage\relax
   \xdef\@gtempa{\write\@auxout{\string
      \newlabel{#1}{{\thesubequation}{\thepage}}}}}\@gtempa
   \if@nobreak \ifvmode\nobreak\fi\fi\fi\@esphack}
\def\subeqnarray{\stepcounter{equation}
\let\@currentlabel=\theequation\global\c@subequation\@ne
\global\@eqnswtrue
\global\@eqcnt\z@\tabskip\@centering\let\\=\@subeqncr
$$\halign to \displaywidth\bgroup\@eqnsel\hskip\@centering
  $\displaystyle\tabskip\z@{##}$&\global\@eqcnt\@ne
  \hskip 2\arraycolsep \hfil${##}$\hfil
  &\global\@eqcnt\tw@ \hskip 2\arraycolsep
  $\displaystyle\tabskip\z@{##}$\hfil
   \tabskip\@centering&\llap{##}\tabskip\z@\cr}
\def\endsubeqnarray{\@@subeqncr\egroup
                     $$\global\@ignoretrue}
\def\@subeqncr{{\ifnum0=`}\fi\@ifstar{\global\@eqpen\@M
    \@ysubeqncr}{\global\@eqpen\interdisplaylinepenalty \@ysubeqncr}}
\def\@ysubeqncr{\@ifnextchar [{\@xsubeqncr}{\@xsubeqncr[\z@]}}
\def\@xsubeqncr[#1]{\ifnum0=`{\fi}\@@subeqncr
   \noalign{\penalty\@eqpen\vskip\jot\vskip #1\relax}}
\def\@@subeqncr{\let\@tempa\relax
    \ifcase\@eqcnt \def\@tempa{& & &}\or \def\@tempa{& &}
      \else \def\@tempa{&}\fi
     \@tempa \if@eqnsw\@subeqnnum\refstepcounter{subequation}\fi
     \global\@eqnswtrue\global\@eqcnt\z@\cr}
\let\@ssubeqncr=\@subeqncr
\@namedef{subeqnarray*}{\def\@subeqncr{\nonumber\@ssubeqncr}\subeqnarray}
\@namedef{endsubeqnarray*}{\global\advance\c@equation\m@ne%
                           \nonumber\endsubeqnarray}

\makeatletter
\@addtoreset{equation}{section}
\makeatother
\renewcommand{\theequation}{\thesection.\arabic{equation}}

\def\dalemb#1#2{{\vbox{\hrule height .#2pt
        \hbox{\vrule width.#2pt height#1pt \kern#1pt
                \vrule width.#2pt}
        \hrule height.#2pt}}}

\let\a=\alpha \let\b=\beta   \let\e=\epsilon
  \let\q=\theta  
 \let\m=\mu \let\n=\nu

\def\nn{\nonumber} \def\bd{\begin{document}} \def\ed{\end{document}}
\def\ds{\documentstyle} \let\fr=\frac \let\bl=\bigl \let\br=\bigr
\let\Br=\Bigr \let\Bl=\Bigl
\let\bm=\bibitem
\let\na=\nabla
\let\pa=\partial \let\ov=\overline
\def\ie{{\it i.e.\ }}
\newcommand{\be}{\begin{equation}}
\newcommand{\ee}{\end{equation}}
\def\ba{\begin{array}}
\def\ea{\end{array}}
\def\ft#1#2{{\textstyle{{\scriptstyle #1}\over {\scriptstyle #2}}}}
\def\fft#1#2{{#1 \over #2}}
\def\del{\partial}
\def\sst#1{{\scriptscriptstyle #1}}
\def\st#1{{\scriptstyle #1}}
\def\oneone{\rlap 1\mkern4mu{\rm l}}
\def\e7{E_{7(+7)}}
\def\td{\tilde}
\def\wtd{\widetilde}
\def\im{{\rm i}}
\def\bog{Bogomol'nyi\ }
\def\q{{\tilde q}}
\def\hast{{\hat\ast}}
\def\0{{\sst{(0)}}}
\def\1{{\sst{(1)}}}
\def\2{{\sst{(2)}}}
\def\3{{\sst{(3)}}}
\def\4{{\sst{(4)}}}
\def\5{{\sst{(5)}}}
\def\6{{\sst{(6)}}}
\def\7{{\sst{(7)}}}
\def\8{{\sst{(8)}}}
\def\n{{\sst{(n)}}}
\def\oo{{\"o}}
\def\hA{\hat{\cal A}}
\def\ns{{\sst {\rm NS}}}
\def\rr{{\sst {\rm RR}}}
\def\tH{{\widetilde H}}
\def\tB{{\widetilde B}}
\def\cA{{\cal A}}
\def\cF{{\cal F}}
\def\tF{{\wtd F}}
\def\Z{\rlap{\sf Z}\mkern3mu{\sf Z}}
\def\ep{{\epsilon}}
\def\IIA{{\rm IIA}}
\def\IIB{{\rm IIB}}
\def\ads{{\rm AdS}}
\def\R{\rlap{\rm I}\mkern3mu{\rm R}}
\def\btheta{{\bar\theta}}
\def\ttheta{{{\tilde\theta}}}
\def\bttheta{{{\bar\ttheta}}}
\def\sw{{(1\leftrightarrow2)}}
\def\bg{{{\bf g}}}

\def\Ei{{\hbox{Ei}}}
\def\Ci{{\hbox{Ci}}}
\def\Si{{\hbox{Si}}}

\newcommand{\ho}[1]{$\, ^{#1}$}
\newcommand{\hoch}[1]{$\, ^{#1}$}
\newcommand{\bea}{\begin{eqnarray}}
\newcommand{\eea}{\end{eqnarray}}
\newcommand{\ra}{\rightarrow}
\newcommand{\lra}{\longrightarrow}
\newcommand{\Lra}{\Leftrightarrow}
\newcommand{\ap}{\alpha^\prime}
\newcommand{\bp}{\tilde \beta^\prime}
\newcommand{\tr}{{\rm tr} }
\newcommand{\Tr}{{\rm Tr} }
\newcommand{\NP}{Nucl. Phys. }
\newcommand{\tamphys}{\it Center for Theoretical Physics,
Texas A\&M University, College Station, TX 77843}
\newcommand{\upenn}{\it Dept. of Phys. and Astro.,
University of Pennsylvania,
Philadelphia, PA 19104}

\newcommand{\auth}{M. Cveti\v{c}\hoch{\dagger + 1}, 
H. L\"u\hoch{\dagger1}, 
C.N. Pope\hoch{\ddagger2} and K.S. Stelle\hoch{\star3} }

\begin{document}
\begin{flushright}
\hfill{
UPR/0852-T \ \   
CTP TAMU-31/99 \ \
SISSA-Ref.\ 88/99/EP\ \  
Imperial/TP/98-99/63}\\
\hfill{
NSF-ITP-99/086\ \ \ \ \ \ \ \ July 1999 \ \ \ \ \ \ \ \ \ 
\bf hep-th/9907202}\\
\end{flushright}


\begin{center}

{\large {\bf T-Duality in the Green-Schwarz Formalism, and the\\
 Massless/Massive IIA Duality Map}}

\vspace{15pt}

\auth

\vspace{7pt}
{\hoch{\dagger}\upenn}

\vspace{7pt}
{\hoch{+}\it Institute for Theor. Phys., Univ. of Santa Barbara, 
Santa Barbara, CA 93106}

\vspace{7pt}
{\hoch{\ddagger}\tamphys}

\vspace{10pt}
{\hoch{\ddagger}
\it SISSA, Via Beirut No. 2-4, 34013 Trieste, Italy}

\vspace{7pt}
{\hoch{\star}{\it The Blackett Laboratory, Imperial College, Prince
Consort Road, London SW7 2BZ. }}

\vspace{15pt}

\underline{ABSTRACT}
\end{center}

    We derive a component-field expansion of the Green-Schwarz action
for the type IIA string, in an arbitrary background of massless NS-NS
and R-R bosonic fields, up to quadratic order in the fermionic
coordinates $\theta$.  Using this action, we extend the usual
derivation of Buscher T-duality rules to include not only NS-NS, but
also R-R fields.  Our implementation of the T-duality transformation
rules makes use of adapted background-field parametrizations, which
provide a more geometrically natural and elegant description for the
duality maps than the ones previously presented.  These T-duality
rules allow us to derive the Green-Schwarz action for the type IIB
string in an arbitrary background of massless NS-NS and R-R bosonic
fields, up to $O(\theta^2)$.  Implemention of another T-duality
transformation on this type IIB action then allows us also to derive
the Green-Schwarz action for the massive IIA string. By further
considering T-duality transformations for backgrounds with the two
$U(1)$ isometries of a 2-torus, we give a string-theoretic derivation
of the direct T-duality relation between the massless and massive type
IIA strings.  In addition, we give an explicit construction of the
$D=8$ $SL(3,\R)\times SL(2,\R)$ invariant supergravity with two mass
parameters that form a doublet under the $SL(2,\R)$ factor.

{\vfill\leftline{}\vfill
\vskip 10pt \footnoterule
{\footnotesize \hoch{1}
Research supported in part by DOE grant DOE-FG02-95ER40893
\vskip  -12pt} \vskip   14pt
{\footnotesize \hoch{2}
Research supported in part by DOE grant DE-FG03-95ER40917
\vskip -12pt} \vskip 14pt
{\footnotesize \hoch{3}
Research supported in part by the EC under TMR
contract ERBFMRX-CT96-0045.
\vskip -12pt}  \vskip  14pt
}

\pagebreak
\setcounter{page}{1}

\tableofcontents
\addtocontents{toc}{\protect\setcounter{tocdepth}{2}}
\newpage

\section{Introduction\label{sec:intro}}

     T-duality is the most securely founded of the duality symmetries
of string theory, being grounded in worldsheet field manipulations
that do not change the corresponding conformal field theory (see
reviews \cite{alvarez2lozano,gpr}). It has been discussed mainly in
the Neveu-Schwarz-Ramond formalism, where couplings to the
Ramond-Ramond sector of superstring theory do not appear in the
classical string action.  However, when one is concerned with the
effects of T-duality on nontrivial R-R backgrounds, it is more
appropriate to use the Green-Schwarz formulation.  In this paper we
develop procedures for deriving T-duality in the Green-Schwarz
formalism.  We first apply the method to a derivation of the standard
type IIA/IIB string T-duality, including R-R backgrounds.  Next, we
extend the discussion to allow for backgrounds associated with the
massive type IIA theory.  Since this extension of the theory
inherently involves R-R backgrounds, the Green-Schwarz formalism is
ideally suited to describing the extended T-duality symmetries that
relate the massive IIA string to the type IIB string.  Finally, we
apply our discussion of T-duality in the Green-Schwarz formalism to
the case of type IIA strings propagating in backgrounds with two
abelian isometries.  By this means we are able to exhibit explicitly
the T-duality relation between the massless and massive IIA theories.

    We begin by obtaining a component-field expansion of the
superfield form of the Green-Schwarz action for the type IIA
superstring in an arbitrary background of bosonic NS-NS and R-R
fields, working to order $O(\theta^2)$ in the fermionic
coordinates. We do this by making a double dimensional reduction
\cite{dhis} starting from the superspace form of the $D=11$
supermembrane action \cite{bst}, and using previous results for the
component expansions of the $D=11$ superfields
\cite{cremmerferrara,brinkhowe,dwpp}.  Thus we obtain the
component-field expansion of the type IIA Green-Schwarz string action
given in (\ref{type2alag}), to $O(\theta^2)$ in the fermionic
coordinates, in arbitrary massless bosonic NS-NS and R-R fields.

     Performing next the T-duality transformation for a single $U(1)$
isometry, we show how an appropriate set of background-field
definitions significantly simplifies the T-duality derived
transformations of these fields.  We do this by adopting variable
choices for the background fields that are geometrically adapted to
the background with its assumed abelian isometry, which is necessary
for implementation of the T-duality map. This leads to a simple set of
expressions for the background-field transformations that precisely
matches the known forms of nonlinear symmetry transformations in the
corresponding effective supergravity theory. Since, in our discussion,
we are using a formalism where all the massless bosonic fields are
non-vanishing, this allows us to derive T-duality transformation rules
for the R-R as well as the NS-NS backgrounds, in a manner directly
comparable with previous field-theoretic derivations, but now instead
within the framework of the string action.

    String-theory T-duality transformations have an advantage over
those derived in the effective field theory, in that they are carried
out without suppressing string excitations along the directions of the
assumed isometries of the background fields. Thus, whereas in
supergravity field theories the assumption of isometries means that
one has effectively made a Kaluza-Klein reduction to a
lower-dimensional theory, this is not the case in string
theory. Accordingly, in string theory, one may access the larger
groups of nonlinear symmetry transformations that appear in
supergravity theories upon dimensional reduction, but without actually
sacrificing the higher-dimensional modes of the string itself. Thus in
our approach we obtain the type IIA/IIB T-duality in the more general
string-theoretic framework, including R-R background fields, while at
the same time employing the more geometrically adapted description
commonly used in field-theoretic discussions.

    The component-field expansion of the usual type IIA Green-Schwarz
string that we obtain includes non-vanishing backgrounds for the
2-form and 4-form R-R field strengths $F_\2$ and $F_\4$.  By
performing a T-duality transformation with a single $U(1)$ isometry,
we show how the action can again be interpreted as a covariant
ten-dimensional action, but now describing the type IIB string.  By
this means we derive for the first time a component-field expansion of
the type IIB Green-Schwarz string action, again to $O(\theta^2)$ in
the fermionic coordinates, in arbitrary massless bosonic backgrounds,
including the 1-form, 3-form and 5-form R-R field strengths $\wtd
F_\1=d\chi$, $\wtd F_\3$ and $\wtd F_\5$.  The action is given by
(\ref{type2baction}). 

    From the type IIB Green-Schwarz action we then perform a further
T-duality transformation, for a single $U(1)$ isometry along a
coordinate $y$, in which the field strength $\wtd F_\1$ is allowed to
have a non-vanishing component in the direction of the isometry, $\wtd
F_\1\longrightarrow \wtd F_\1 + m\, dy$.  From a field-theoretic
standpoint this corresponds to performing a Scherk-Schwarz generalised
reduction, where the axion $\chi$ is allowed a linear dependence on
the compactification coordinate $y$.  As is well known, this leads to
a massive supergravity in $D=9$ which is identical, after a field
transformation, to the ordinary dimensional reduction of massive 
\cite{romans} type IIA
supergravity \cite{bdgpt} (see also \cite{ortin2,llps}).  In our
discussion of T-duality at the level of the superstring action, we are
able to derive the additional 0-form contribution of the mass term $m$
to the background of the IIA theory, thereby providing an explicit
expression for the Green-Schwarz action for the massive IIA string,
given by (\ref{type2amassive}).

    Since the type IIB and the massive type IIA theories are related
by an appropriately-generalised T-duality on a circle, and on the
other hand the massless IIA theory is related to the IIB theory by a
standard T-duality on a circle, one can expect that it should be
possible to derive a direct T-duality relation between the massless
and massive IIA theories themselves, in which the background fields
are assumed to have the two $U(1)$ isometries of a 2-torus.  Indeed,
in \cite{clpst} a massive supergravity was obtained in $D=8$, by
performing a generalised Scherk-Schwarz reduction of massless type IIA
supergravity.  Although not manifestly the same as the theory that one
obtains by reducing massive type IIA supergravity on the 2-torus, it
is in fact equivalent up to field redefinitions in $D=8$.  This formed
an ingredient in the discussion in \cite{hull}, where the notion of
massless/massive IIA duality was developed.  The arguments presented
there involved the comparison of D8-brane and D6-brane solutions of
the supergravities in $D=8$.  In the present paper, we are able to
present a direct and explicit formulation of massless/massive IIA
duality, at the level of the Green-Schwarz string action, showing
precisely how the T-duality mapping between the two theories is
implemented.

\section{Type IIA Superstring Action from the Supermembrane}

    We start from the worldvolume action for the supermembrane in $D=11$.  
This is given by \cite{bst}
\be
I_3 = \int d^3\xi \Big( -\sqrt{-\hat h} -\ft16 \ep^{\hat i\hat j\hat k}\, 
\del_{\hat i} Z^M\, 
\del_{\hat j} Z^N\, \del_{\hat k} Z^P\, \hat A_{PNM}\Big) 
\,,\label{d11mem}
\ee
where $\hat h_{\hat i \hat j} = \del_{\hat i}Z^M\, \hat E^{\hat m}_M
\, \del_{\hat j} Z^N\, \hat E^{\hat n}_N\, \eta_{\hat m\hat n}$ 
is the world-volume metric,
$Z^M=(X^{\hat\mu}, \theta^\alpha)$ denotes the eleven bosonic and 32
fermionic spacetime coordinates, and $\hat A_{PNM}$ is the 3-form
superfield.  The supervielbein $\hat E^{A}_{M}$ and 3-form $\hat A$
were obtained to leading order in $\theta$ in
\cite{cremmerferrara,brinkhowe}, and more recently they were
completely calculated to $O(\theta^2)$ \cite{dwpp}.  To this order,
setting the spinor background fields to zero, they are given by:
\bea
\hat E^{\hat m}_{\hat\mu} &=& \hat e^{\hat m}_{\hat\mu} + \im\, \btheta\,
\Gamma^{\hat m}(\ft14 \hat\omega_{\hat\mu}{}^{\hat n\hat p}\, 
\Gamma^{\hat n\hat p} + \hat T_{\hat\mu}{}^{\hat n\hat p\hat q\hat r}
\, \hat F_{\hat n\hat p\hat q\hat r})\theta\,,\nn\\
\hat E^{a}_{\hat\mu} &=& \ft14
\hat\omega_{\hat\mu}{}^{\hat m\hat n}\, 
(\Gamma_{\hat m\hat n}\theta)^a + (\hat T_{\hat\mu}{}^{\hat m\hat
n\hat p \hat q}\,\theta)^a\, \hat F_{\hat m\hat n\hat p\hat q}\,,\nn\\
\hat E^{\hat m}_\a &=& -\im\, (\btheta\, \Gamma^{\hat m})_{\a}\,, \qquad
\hat E^a_\a = \delta^a_\a + M^a_\a\,,\label{d11viel}\\
\hat A_{\hat\mu\hat\nu\hat\rho} &=& A_{\hat\mu\hat\nu\hat\rho} -
3\im\, \btheta\, \hat\Gamma_{[\hat\mu\hat\nu} \, \Big( \ft14
\hat\omega_{\hat\rho]}{}^{\hat m\hat n}\, \Gamma_{\hat m\hat n} + \hat
T_{\hat\rho]}{}^{ \hat m\hat n\hat p\hat q}\, \hat F_{\hat m\hat n\hat
p\hat q}\Big)\theta\,,\nn\\
\hat A_{\hat\mu\hat\nu \a} &=& \im\, (\btheta\, \hat\Gamma_{\hat\mu
\hat\nu})_\a \,,\qquad \hat A_{\hat\mu\a\b} = -(\btheta\hat
\Gamma_{\hat\mu \hat \nu})_{(\a}\, (\btheta\, \hat\Gamma^{\hat\nu}
)_{\b)}\,,\nn\\
\hat A_{\a\b\gamma} &=& -\im\,  (\btheta\hat 
\Gamma_{\hat\mu \hat \nu})_{(\a}\,
 (\btheta\, \hat\Gamma^{\hat\mu} )_{\b}\,  (\btheta\, \hat\Gamma^{\hat\nu}
)_{\gamma)}\,.\nn
\eea
The notation here is as follows.  Hatted indices $\hat\mu$ are used
for eleven-dimensional bosonic coordinates and hatted indices $\hat m$
are used for the eleven-dimensional bosonic tangent-space.  Fermionic
coordinate and tangent-space indices are denoted by $\a$ and $a$
respectively.  The quantity $\hat
T_{\hat\mu}{}^{\hat\nu\hat\rho\hat\sigma\hat\lambda}$ is given by
\be
\hat T_{\hat\mu}{}^{\hat\nu\hat\rho\hat\sigma\hat\lambda} = 
\ft1{288}(\hat\Gamma_{\hat\mu}{}^{\hat\nu\hat\rho\hat\sigma\hat\lambda} -
8 \delta_{\hat\mu}^{[\hat\nu}\, \hat\Gamma^{\hat\rho\hat 
\sigma \hat \lambda]})\,.
\ee
Whenever coordinate indices appear on the Dirac matrices, we place
hats on the matrices to indicate that the eleven-dimensional bosonic
vielbein $\hat e^{\hat m}_{\hat\mu}$ has been used.  The quantity
$M^a_\a$ appearing in (\ref{d11viel}) is of order $\theta^2$, but
does not contribute to any results at this order.\footnote{Note that
we have changed from the spinor-conjugation convention used in 
\cite{dwpp}, where $\bar\psi=\im\, \psi^\dagger\, \Gamma^t$, to the
more familiar one where  $\bar\psi=\psi^\dagger\, \Gamma^t$.  See 
the Addendum section \ref{addsec} for a more detailed discussion.}

    We now perform a double-dimensional reduction from $D=11$, in
order to obtain the superspace action for the type IIA string in
$D=10$, expanded to order $\theta^2$, in the presence of bosonic
background fields.  The bosonic coordinates $X^{\hat\mu}$ are split as
$X^{\hat\mu} = (X^\mu, y)$, and the eleventh spacetime coordinate $y$
is set equal to the third world-volume coordinate $\xi^3$;
$y=\xi^3$. Otherwise, all spacetime coordinates are taken to be
independent of $\xi^3$, and the background fields are taken to be
independent of $y$.  The reduction ans\"atze for these bosonic
background fields are the usual ones, namely
\bea
d\hat s^2_{11} &=& e^{-\fft23\phi}\ ds^2_{10} + e^{\fft43\phi}\, (dy +
A_\1)^2\,,\label{metred}\\
\hat A_\3 &=& A_\3 + A_\2\wedge dy\,,\label{3formred} 
\eea
where $d\hat s^2_{11}= \hat g_{\hat\mu\hat\nu} dX^{\hat\mu}\,
dX^{\hat\nu}$ and $ds^2_{10}= g_{\mu\nu}\, dX^\mu\, dX^\nu$, with the
metrics given by $\hat g_{\hat\mu\hat\nu} = \hat e^{\hat
m}_{\hat\mu}\, \hat e^{\hat n}_{\hat\nu}\, \eta_{\hat m\hat n}$ and
$g_{\mu\nu} = e^m_\mu\, e^n_\nu\, \eta_{mn}$ respectively.  Note that
the ten-dimensional metric is in the string frame here.  We also
choose to make a rescaling of the fermionic coordinates, in the
process of reducing from $D=11$ to $D=10$, namely,
$\theta\longrightarrow e^{-\fft16\phi}\, \theta$.  This ensures that
in $D=10$, the supercoordinate transformations $\delta X^\mu =\im\, 
\btheta\, \Gamma^\mu\, \epsilon$ and $\delta\theta=\epsilon$ will take
their canonical forms.

    In order to derive the ten-dimensional type IIA superstring action
from the eleven-dimensional supermembrane action (\ref{d11mem}) using
this reduction scheme, one can follow either of two procedures. One
may first obtain the ten-dimensional supervielbein components by
dimensional reduction in superspace, and then substitute them into the
superspace version of the $D=10$ type IIA superstring action,
\be
I_2 = \int d^2\xi \Big( -\sqrt{-h} -\ft12 \ep^{ij}\, 
\del_{\hat i} Z^M\, 
\del_{\hat j} Z^N\, A_{NM}\Big) 
\,,\label{d10str}
\ee
or alternatively one may start directly from the $D=11$ action
(\ref{d11mem}) and work out the reduction directly in the action at
the component-field level. In the first procedure, one has to be
careful to apply a local Lorentz transformation in superspace in order
to put the $D=11$ vielbein into Kaluza-Klein reduction form, with
$\hat E_y^m=0$, and one also needs to perform a superspace Weyl
rescaling \cite{dhis,aks} $(\hat E_\mu^m=\Phi^{-\fft13}E_\mu^m,\, \hat
E_\alpha^m=\Phi^{-\fft13}e^{\fft16\phi}E_\alpha^m)$, where
$\Phi^{\fft23} = \hat E_y^{11}$.  It is also convenient to perform
additional ten-dimensional local Lorentz transformations to simplify
the result. For the vielbein components needed in
the type IIA string action, we then obtain after some algebra the
following
\bea 
E_\mu^m &=& e^m_\mu + \ft{\im}4 \omega_\mu^{pq}\, \bar\theta \, \Gamma^m\, 
\Gamma_{pq}\, \theta -\ft{\im}8 \bar\theta\, \Gamma_{11}\, 
\Gamma^m\, \Gamma^{pq}\, 
\theta\, F_{\mu pq} + \ft{\im}{16} e^\phi\, \bar\theta\, \Gamma_{11}\, 
\Gamma^m\, \Gamma^{pq}\, \Gamma_\mu\, \theta\, F_{pq}\nn\\
&& 
+\ft{\im}{192} e^\phi\, \bar\theta\, \Gamma^m\, \Gamma^{p_1\cdots p_4}\, 
\Gamma_\mu\, \theta\, F_{p_1\cdots p_4}\,,\nn\\
E_\alpha^m &=& -\im\, (\bar\theta\Gamma^m)_\alpha\,.\label{d10vielbeins}
\eea 
The type IIA superstring action may then be obtained by inserting
these expressions, together with the corresponding reduction of the
3-form field $A$, into the $D=10$ superspace form (\ref{d10str}) of
the string action. Alternatively, one may apply the reduction scheme
directly in the eleven-dimensional action (\ref{d11mem}). In either
way we find, after some lengthy algebra, that the ten-dimensional type
IIA string action, for arbitrary bosonic background fields and
expanded to $O(\theta^2)$, is given by $I_2 =\int d^2\xi\, {\cal
L}_2$, where\footnote{A $\Gamma_{11}$ factor was accidentally omitted in
the $F_{\nu\rho\sigma}$ terms in 
the earlier and published versions of this paper.  See the Addendum 
section \ref{addsec} for a detailed discussion of corrections and 
changes of convention and notation.}
\bea
{\cal L}_2 &=& -\ft12 \sqrt{-h}\, h^{ij}\, \del_i X^\mu\, \del_j\,
X^\nu\, g_{\mu\nu} + \ft12\ep^{ij}\, \del_i\, X^\mu\, \del_j\, X^\nu\,
A_{\mu\nu} \nn\\
&&- \im\, \btheta(\sqrt{-h}\, h^{ij} -\ep^{ij}\, \Gamma_{11})\,
\Gamma_\mu\, D_j\theta\, \del_i X^\mu \nn\\
&&+\ft{\im}8 
\del_i X^\mu\, \del_j X^\nu\, \btheta(\sqrt{-h}\, h^{ij}
-\ep^{ij}\, \Gamma_{11})\, \Gamma_{11}\Gamma_\mu{}^{\rho\sigma}\, \theta\,
F_{\nu\rho \sigma}  \label{type2alag}\\
&&-\ft{\im}{16} \del_i X^\mu \del_j X^\nu e^{\phi}\, \btheta 
(\sqrt{-h}\, h^{ij} -\ep^{ij}\,\Gamma_{11})\, \Big(\Gamma_{11}\,
\Gamma_\mu\, \Gamma^{\rho\sigma}\, \Gamma_\nu\, F_{\rho\sigma} + \ft1{12}
\Gamma_\mu\, \Gamma^{\rho\sigma\lambda\tau}\, \Gamma_\nu\,
F_{\rho\sigma\lambda\tau} \Big)\, \theta\,,\nn
\eea
where
\be
D_i\theta \equiv \del_i\theta + \ft14 \del_i X^\mu\,
\omega_\mu{}^{mn}\, \Gamma_{mn}\, \theta\,.
\ee
The field strengths are given by
\be
F_\4 = dA_\3 - A_\1\wedge dA_\2\,,\qquad F_\3= dA_\2\,,\qquad 
F_\2 = dA_\1\,.
\ee
For later convenience, the gamma matrix $\Gamma_{11}$ in
(\ref{d10vielbeins}, \ref{type2alag}) is taken to be the {\it
negative} of the matrix $\Gamma_{\hat m}$ that corresponds to setting
$\hat m$ equal to its eleventh value.  $\Gamma_{11}$ is then the
chirality operator in $D=10$.\footnote{Our conventions are as follows.
The Dirac matrices satisfy $\{\Gamma_{\hat m}, \Gamma_{\hat n}\}= 2
\eta_{\hat m\hat n}$, with $\eta={\rm diag}\, (-1,+1,\ldots +1)$.  For
a Majorana spinor $\theta$, the expressions $\btheta \Gamma_{\hat
m_1\cdots \hat m_p}\, \theta$ vanish for $p=(1,2,5,6,9,10)$ and are
non-vanishing for $p=(0,3,4,7,8,11)$.}

   A number of remarks on this result are now in order.  If we
consider a background in which only the metric and 2-form potential
$A_{\mu\nu}$ are excited, the top two lines in (\ref{type2alag}) are 
precisely of the standard form, up to $O(\theta^2)$.  The third line
describes an additional $O(\theta^2)$ coupling of the NS-NS field
strength $F_\3$, over and above the standard bosonic coupling to the
worldsheet. The final line in (\ref{type2alag}) describes the
couplings\footnote{The odd-parity R-R terms proportional to $\ep^{ij}$
were presented in \cite{tseytlin}; they are in broad structural
agreement with ours.} of the R-R background
fields $F_\2$ and $F_\4$. The factor of $e^\phi$ that multiplies the
R-R terms in (\ref{type2alag}) implies that $\theta$ loop diagrams
involving two insertions of the R-R vertices will give rise to the
expected additional $e^{2\phi}$ factor in the couplings
of the kinetic terms of the R-R fields in the low-energy effective action.

    One further remark that is appropriate at this stage is to note
that there is no $\sqrt{-h}\, R_\2\,\phi$ coupling of the dilaton to
the worldsheet.  This is what one should expect; such a term would
explicitly break the conformal invariance of the classical world-sheet
action (\ref{type2alag}), which would contradict the manifest
conformal invariance of the classical reduction procedure (see
\cite{aks}).  Moreover, a $\sqrt{-h}\, R_\2\,\phi$ term would also
break the manifest spacetime supersymmetry of the Green-Schwarz
formalism.  Whether such a term is actually needed in this formalism
remains an open question. In order to calculate conformal anomalies in
the Green-Schwarz formalism, one must carefully fix the
$\kappa$-symmetry gauge and handle its infinitely recursive ghost
sector. In the explicit $\beta$-function calculations that have been
performed, for the Green-Schwarz heterotic string at the one and
two-loop levels \cite{gnz}, no need has been found for a $\sqrt{-h}\,
R_\2\,\phi$ term. It may be that such a term only appears out of the
quantum measure upon making the variable changes needed for a
transition from the Neveu-Schwarz-Ramond to the Green-Schwarz
formalism.

\section{T-duality and the Type IIB Superstring Action}

\subsection{T-duality in the Green-Schwarz action}

    T-duality in the NS-NS sector has been extensively studied at the
level of the string sigma model, starting with \cite{buscher}.  For
the type I string, the resulting ``Buscher rules'' correspond in $D=9$
to a discrete $Z_2$ symmetry.  For type IIA and IIB strings, whose
NS-NS sectors coincide but whose R-R sectors differ, it corresponds to
a transformation that maps back and forth between the two theories.
In an NSR type of sigma model, one does not see the couplings of the
R-R background fields to the worldsheet, and so the methods used in
\cite{buscher} are not directly applicable to the complete type II
theories.  In fact until now, the analogues of the Buscher rules for
the R-R fields have been derived only at the level of the effective
low-energy field theory, by performing standard Kaluza-Klein
reductions of the type IIA and IIB supergravities and making the
necessary identifications of fields that relate the two theories in
$D=9$, as in \cite{bho,bdgpt}.  In this section, we shall derive the
generalisation of the Buscher rules for the Green-Schwarz superstring
actions.  Since, as we have presented in the previous section, these
actions include the couplings of the R-R fields, this will allow us to
obtain a worldsheet derivation of the ``Buscher rules'' for the R-R
fields.

   To begin, we write down a generic worldsheet Lagrangian with the
general structure of (\ref{type2alag}):
\bea
{\cal L}&=& -\ft12 \sqrt{-h}\, h^{ij}\, \del_i X^\mu\,  \del_j X^\nu \, 
G_{\mu\nu} + \ft12 \ep^{ij}\, \del_i X^\mu\, \del_j\, X^\nu\,
B_{\mu\nu} \nn\\
&&+\sqrt{-h}\, h^{ij}\, \del_i X^\mu\, {\cal G}_{j\mu} + \ep^{ij}\,
\del_i X^\mu\, {\cal B}_{j\mu}\,.\label{genlag}
\eea
Note that by comparing this general form with (\ref{type2alag}), the
quantities $G_{\mu\nu}$ and $B_{\mu\nu}$ will contain not only the
usual $\theta$-independent terms $g_{\mu\nu}$ and $A_{\mu\nu}$, but
also certain $\theta^2$ terms as well.  In fact, it will prove useful
to write these as
\be
G_{\mu\nu} = \bg_{\mu\nu} + Q_{\mu\nu}\,,\qquad
B_{\mu\nu} = A_{\mu\nu} + P_{\mu\nu}\,,\label{pqdef}
\ee
where $Q_{\mu\nu}=Q_{(\mu\nu)}$ and $P_{\mu\nu}=P_{[\mu\nu]}$
represent the symmetric and antisymmetric $O(\theta^2)$ terms coupling
to $\del_i X^\mu \, \del_j X^\nu$.  The quantities ${\cal G}_{j\mu}$
and ${\cal B}_{j\mu}$, which couple to the single $\del_i X^\mu$
pullback, will correspond to the $\btheta\, \del_i\theta$ parts of the
fermionic kinetic terms in (\ref{type2alag}).

    We now follow the analogue of the standard procedure developed for
the bosonic string in \cite{buscher}.  Thus we suppose that there is an
abelian isometry of the ten-dimensional background fields, such that
in an adapted coordinate system the fields are all independent of the
tenth of the coordinates $X^\mu$.  We shall accordingly decompose the
coordinates as $X^\mu = (X^{\mu'}, Y)$.  The next step is to replace
$\del_i Y$ in the Lagrangian (\ref{genlag}) by a worldsheet vector
$V_i$, at the same time introducing a Lagrange multiplier $Z$ (a
worldsheet scalar) and adding the term $\ep^{ij}\, \del_i Z\, V_j$.
The equation of motion for $Z$ enforces the irrotationality of $V_i$.
By instead varying the action with respect to $V_i$, and substituting
its algebraic solution back into the action, we obtain
\bea
{\cal L}&=& -\ft12 \sqrt{-h}\, h^{ij}\, \del_i X^{\mu'}\,  
\del_j X^{\nu'} \, 
G_{\mu'\nu'} + \ft12 \ep^{ij}\, \del_i X^{\mu'}\, \del_j\, X^{\nu'}\,
B_{\mu'\nu'} \nn\\
&&+\sqrt{-h}\, h^{ij}\, \del_i X^{\mu'}\, {\cal G}_{j\mu'} + \ep^{ij}\,
\del_i X^{\mu'}\, {\cal B}_{j\mu'}\nn\\
&&-\ft12\sqrt{-h}\, h^{ij}\, G_{00}^{-1}\, (u_i\, u_j -v_i\, v_j) +
\ep^{ij}\, G_{00}^{-1}\, u_i\, v_j\,,\label{genlag2}
\eea
where
\be
u_i \equiv \del_i Z + \del_i X^{\mu'}\, B_{\mu' 0} - {\cal B}_{i0}\,,
\qquad v_i\equiv -\del_i X^{\mu'}\, G_{\mu' 0} + {\cal G}_{i0}\,.
\ee
Note that, following the conventional notation, we associate the index
``0'' with the $Y$ coordinate.  

     The Lagrangian (\ref{genlag2}) can be seen to be of the same
general form as (\ref{genlag}), with $Z$ rather than $Y$ now viewed as
the tenth coordinate.  Recasting (\ref{genlag2}) in the
form (\ref{genlag}) implies that there will now be a transformed set
of background fields, which we shall denote by $\wtd G_{\mu\nu}$,
$\wtd B_{\mu\nu}$, $\wtd{\cal G}_{j\mu}$ and $\wtd{\cal B}_{j\mu}$.
Recalling that $G_{\mu\nu}$ and $B_{\mu\nu}$ include terms
both of order $\theta^0$ and $\theta^2$, as indicated in
(\ref{pqdef}), we may then read off the full set of transformed
background fields, correct to order $\theta^2$.  At zeroth order in
$\theta$ we find the usual Buscher rules,
\bea
\tilde \bg_{\mu'\nu'} &=& \bg_{\mu'\nu'} + \bg_{00}^{-1}\, (A_{\mu' 0}\,
A_{\nu' 0} - \bg_{\mu'0}\, \bg_{\nu'0})\,,\nn\\
\td \bg_{\mu'0} &=& \bg_{00}^{-1}\, A_{\mu'0}\,,\qquad 
\td \bg_{00}= \bg_{00}^{-1}\,,\label{bosbus}\\
\wtd A_{\mu'\nu'} &=& A_{\mu'\nu'} + \bg_{00}^{-1}\, (\bg_{\mu'0}\,
A_{\nu'0} - \bg_{\nu'0}\, A_{\mu'0})\,,\nn\\
\wtd A_{\mu'0} &=& \bg_{00}^{-1}\, \bg_{\mu'0}\,.\nn
\eea
At $O(\theta^2)$, we obtain the following rules:
\bea
\wtd Q_{\mu'\nu'} &=& Q_{\mu'\nu'} +2\bg_{00}^{-1}\, (P_{(\mu'|0|}\,
A_{\nu')0} - Q_{(\mu'|0|}\, \bg_{\nu')0})
+ \bg_{00}^{-2}\, (\bg_{\mu'0}\, \bg_{\nu'0}- A_{\mu' 0}\,
A_{\nu' 0})\, Q_{00}\,,\nn\\
\wtd Q_{\mu'0} &=& \bg_{00}^{-1}\, P_{\mu'0} - \bg_{00}^{-2}\, A_{\mu'0}\,
Q_{00}\,,\qquad \wtd Q_{00} = -\bg_{00}^{-2}\, Q_{00}\,,\nn\\
\wtd P_{\mu'\nu'} &=& P_{\mu'\nu'} - 2 \bg_{00}^{-1}\, ( P_{[\mu'|0|}\,
\bg_{\nu']0} -  Q_{[\mu'|0|}\, A_{\nu']0}) + 2 \bg_{00}^{-2}\, Q_{00}\,
A_{[\mu'|0|}\, \bg_{\nu']0}\,,\nn\\
\wtd P_{\mu'0} &=& \bg_{00}^{-1}\, Q_{\mu'0} - \bg_{00}^{-2}\, \bg_{\mu'0}\,
Q_{00}\,,\label{t2rules}\\
\wtd{\cal G}_{i\mu'} &=& {\cal G}_{i\mu'} + \bg_{00}^{-1}\, ({\cal
B}_{i0}\, A_{\mu'0} - {\cal G}_{i0}\, \bg_{\mu'0})\,, \qquad
\wtd {\cal G}_{i0} = \bg_{00}^{-1}\, {\cal B}_{i0}\,,\nn\\
\wtd{\cal B}_{i\mu'} &=& {\cal B}_{i\mu'} + \bg_{00}^{-1}\, ({\cal
G}_{i0}\, A_{\mu'0} - {\cal B}_{i0}\, \bg_{\mu'0})\,, \qquad
\wtd {\cal B}_{i0} = \bg_{00}^{-1}\, {\cal G}_{i0}\,,\nn
\eea

   In these formulae we have followed tradition \cite{buscher}, in
denoting the components of the metric (the $\theta$-independent terms
in $G_{\mu'\nu'}$, $G_{\mu'0}$ and $G_{00}$) simply as
$\bg_{\mu'\nu'}$, $\bg_{\mu'0}$ and $\bg_{00}$.  There is, however, a
much more natural way to parametrize the ten-dimensional metric in
terms of nine-dimensional fields, namely by using the standard
Kaluza-Klein decomposition from $D=10$ to $D=9$.  Thus the natural
metric to consider in $D=9$ is not $\bg_{\mu'\nu'}$, but rather,
$g_{\mu'\nu'} = \bg_{\mu'\nu'} + \bg_{00}^{-1}\, \bg_{\mu'0}\,
\bg_{\nu'0}$.  (For the mixed and the internal components, there is no
need to make a distinction between $\bg$ and $g$.)  This can be seen
easily from the form of the string-frame Kaluza-Klein reduction
ansatz, which is
\be
d\hat s^2 = ds^2 +  e^{2\lambda}\, (dz+ {\cal A}_\1)^2\,,
\label{kkdecomp}
\ee
where $ds^2= g_{\mu'\nu'}\, dX^{\mu'}\, dX^{\nu'}$, and ${\cal A}_\1=
{\cal A}_{\mu'}\, dX^{\mu'}$ is the Kaluza-Klein vector.  In the type
IIA theory, we have $\lambda=\fft14\phi+\fft{\sqrt7}{4}\varphi$.  In
terms of the metric decomposition (\ref{kkdecomp}), it is easy to see
that the standard bosonic Buscher rules (\ref{bosbus}) can be
rewritten in the more elegant form
\bea
\tilde g_{\mu'\nu'} &=& g_{\mu'\nu'}\,,\qquad 
\wtd{\cal A}_\1 = A_\1\,, \qquad \td\lambda
=-\lambda \,,\nn\\
\wtd A_\2 &=& A_\2 + \cA_\1\wedge A_\1\,,\qquad \wtd A_\1 =\cA_\1\,,
\label{bush1}
\eea 
where the 2-form potential in $D=10$ is reduced to $D=9$ according to
the standard Kaluza-Klein prescription $A_\2\longrightarrow A_\2 +
A_\1\wedge dz$.  Note that the expressions in (\ref{bush1}) are
identical to those that one finds at the field-theory level, when
mapping the dimensionally-reduced type IIA and IIB supergravities into
one another in $D=9$.  (See, for example, \cite{llps}.)

  For the $O(\theta^2)$ terms (\ref{t2rules}), we can also obtain a more 
elegant form by using the technique, familiar in
Kaluza-Klein reductions, of using tangent-space rather than coordinate
indices.  Upon doing so, we find that (\ref{t2rules}) can be
re-expressed in the considerably simpler form
\bea
\wtd Q_{m'n'} &=& Q_{m'n'}\,,\qquad \wtd Q_{m'0} = P_{m'0}\,,\qquad 
\wtd Q_{00} = - Q_{00}\,,\nn\\
\wtd P_{m'n'} &=& P_{m'n'}\,,\qquad \wtd P_{m'0} = Q_{m'0}\,,
\label{bush2}\\
\wtd{\cal G}_{im'} &=& {\cal G}_{im'}\,,\qquad \wtd{\cal G}_{i0} ={\cal
B}_{i0}\,,\nn\\
\wtd {\cal B}_{im'} &=& {\cal B}_{im'}\,,\qquad 
\wtd{\cal B}_{i0} = {\cal G}_{i0}\,,\nn
\eea
where here, $m'$ and $0$ now denote tangent-space indices.  Henceforth, we
shall always use tangent-space indices on $Q_{mn}$ and $P_{mn}$.
Note that all the type IIA and IIB field strengths that appear in our
results, including all their non-linear corrections, will now be precisely 
the same as one finds in the corresponding supergravity theories.  In
the conventions that we are using in this paper, their precise expressions
may all be found in \cite{cjlp1}.

   In order to discuss T-duality in the $O(\theta^2)$ sector, we
must make a nine-dimensional decomposition of the spinor coordinates.
The spinor $\theta$ in $D=10$ has 32 components, split into 16 chiral
and 16 antichiral components.  In $D=9$, we therefore obtain two
16-component spinors.  We shall decompose the ten-dimensional Dirac
matrices as follows:
\bea
\Gamma_{m'} = \pmatrix{0&\gamma_{m'}\cr \gamma_{m'} & 0}\,,\qquad
\Gamma_{0} = \pmatrix{0&\im\cr-\im&0}\,,\qquad
\Gamma_{11} = \pmatrix{\oneone&0\cr 0& -\oneone}\,,
\eea
where $m'$ is a tangent-space index running over the nine-dimensional
range, with corresponding $16\times16$ Dirac matrices $\gamma_{m'}$.
It should be emphasised that $\Gamma_0$ here refers to the Dirac
matrix of the tenth direction (and not the time direction!).  The
32-component spinor $\theta$ of the type IIA theory (\ref{type2alag})
then takes the form $\theta = \theta^{\sst 1} + \theta^{\sst 2}$,
where
\be
\theta^{\sst 1} = \pmatrix{\vartheta^{\sst 1}\cr 0}\,,\qquad 
\theta^{\sst 2} = \pmatrix{0\cr \vartheta^{\sst 2}}\,.
\ee
Thus we have $\Gamma_{11}\theta^{\sst 1} = \theta^{\sst 1}$ and
$\Gamma_{11} \theta^{\sst 2} = -\theta^{\sst 2}$.  

    In the type IIB theory, which we shall obtain by performing a
T-duality transformation, there are two Majorana spinors of positive
chirality in $D=10$, which we shall denote by $\ttheta^{\sst A}$, with
${\st A}=1,2$. These will be related to the spinors $\theta^{\sst A}$
of the type IIA theory by\footnote{The definition of $\td\theta^2$
that we are now using differs from that used in the version of this
paper published in Nucl. Phys. {\bf B573} (2000) 149, and in v1. of
hep-th/9907202, where we defined $\td\theta^2=-\im\, \Gamma_0\,
\theta^2$.  While there was nothing wrong with taking that definition,
it did mean that the two chiral spinors $\td\theta^1$ and
$\td\theta^2$ in the type IIB theory were Majorana and anti-Majorana
respectively.  See the Addendum section \ref{addsec} for a detailed
discussion of our convention changes in this current version.}
\be
\ttheta^{\sst 1} = \theta^{\sst 1}\,,\qquad
\ttheta^{\sst 2} = \Gamma_0\, \theta^{\sst 2}\,.\label{2ato2b}
\ee

\subsection{Type IIA/IIB T-duality}

     With these preliminaries, we are now ready to implement the
T-duality transformation on the type IIA action (\ref{type2alag}).
The new part of our construction, which goes beyond previous results,
involves the consideration of the $O(\theta^2)$ terms, and, in
particular, the R-R field couplings.  Firstly, we read off from
(\ref{type2alag}) the various $O(\theta^2)$ terms in (\ref{genlag}).
In order to bring out the parallelism between the type IIA and type
IIB theories, we express the $\theta$ coordinates of the IIA theory in
the notation $\theta^{\sst A}$ introduced above.  Thus for the terms
coupling to NS-NS background fields we have
\bea
Q^{\ns}_{mn} &=& \ft{\im}4 
[\btheta^{\sst1}\Gamma_{(m}{}^{pq}\, \theta^{\sst1}- 
 \btheta^{\sst2}\Gamma_{(m}{}^{pq}\, \theta^{\sst2}]\,
F^\ns_{n)pq}  +\ft{\im}2 [\btheta^{\sst1} \Gamma_{(m}\,
\Gamma_{|pq|}\, \theta^{\sst1} + 
\btheta^{\sst2} \Gamma_{(m}\,\Gamma_{|pq|}\, \theta^{\sst2}
]\, \omega_{n)}{}^{pq}\,,\nn\\
P^{\ns}_{mn} &=& - \ft{\im}4
[\btheta^{\sst1}\Gamma_{(m}{}^{pq}\, \theta^{\sst1} + 
\btheta^{\sst2}\Gamma_{(m}{}^{pq}\, \theta^{\sst2}]\,
F^\ns_{n)pq} -\ft{\im}2 [\btheta^{\sst1} \Gamma_{(m}\,
\Gamma_{|pq|}\, \theta^{\sst1} - 
\btheta^{\sst2} \Gamma_{(m}\,\Gamma_{|pq|}\, \theta^{\sst2}
]\, \omega_{n)}{}^{pq}\,,\nn\\
{\cal G}_{i m} &=& -\im\, [\btheta^{\sst1}\Gamma_m\, 
\del_i\theta^{\sst 1} +
\btheta^{\sst2}\Gamma_m\, \del_i\theta^{\sst 2}
]\,,\label{nspq}\\
{\cal B}_{im} &=& -\im\, 
[\btheta^{\sst1}\Gamma_m\, \del_i\theta^{\sst 1} -
\btheta^{\sst2}\Gamma_m\, \del_i\theta^{\sst 2}]\,.\nn
\eea
(Note that $\omega_n{}^{pq}$ denotes the
tangent-space components of the spin connection; \ie $\omega^{pq} =
\omega_n{}^{pq}\, e^n$.)  Similarly, we can read off from
(\ref{type2alag}) the contributions
to $Q_{mn}$ and $P_{mn}$ coming from the R-R sector:
\bea
Q^\2_{mn} &=& -\ft{im}4 e^{\phi}\, \btheta^{[{\sst 1}}\,
\Gamma_{(m}\, \Gamma^{pq}\, \Gamma_{n)}\, 
\theta^{{\sst 2}]}\, F_{pq}\,,\nn\\
P^\2_{mn} &=& \ft{\im}4 e^{\phi}\, \btheta^{({\sst 1}}\,
\Gamma_{[m}\, \Gamma^{pq}\, \Gamma_{n]}\, 
\theta^{{\sst 2})}\, F_{pq}\,,\nn\\
Q^\4_{mn} &=& \ft{\im}{48} e^{\phi}\, \btheta^{({\sst 1}}\,
\Gamma_{(m}\, \Gamma^{pqrs}\, \Gamma_{n)}\, \theta^{{\sst 2})}\, F_{pqrs}\,,
\label{rrpq}\\
P^\4_{mn} &=& -\ft{\im}{48} e^{\phi}\, \btheta^{[{\sst 1}}\,
\Gamma_{[m}\, \Gamma^{pqrs}\, \Gamma_{n]}\, 
\theta^{{\sst 2}]}\, F_{pqrs}\,,\nn
\eea
where the superscripts on the $Q_{mn}$ and $P_{mn}$ indicate the
degrees of the R-R field strengths involved.  The complete expressions
for $Q_{mn}$ and $P_{mn}$ are obtained by summing the various NS-NS
and R-R contributions listed in (\ref{nspq}) and (\ref{rrpq}) above.

    We now implement the T-duality transformations for the
$O(\theta^2)$ backgrounds, as given in (\ref{bush2}).  Specifically,
this means that we should take the expressions given by (\ref{nspq})
and (\ref{rrpq}), apply the transformations (\ref{bush2}), and then
seek to re-interpret the resulting tilded quantities as the $9+1$
decompositions of fully ten-dimensional covariant tilded backgrounds
$\wtd P_{mn}$ and $\wtd Q_{mn}$.  It is helpful to study the R-R
terms, given in (\ref{rrpq}), first.  It is straightforward to see
that the last step, of re-interpreting the transformed
nine-dimensional backgrounds as the reductions of covariant
ten-dimensional ones, is impossible if one tries still to use the
original two $\theta^{\sst A}$ fermionic coordinates of the type IIA
action (\ref{type2alag}).  However, if one uses instead the
$\tilde\theta^{\sst A}$ fermionic coordinates, then the transformed
nine-dimensional backgrounds can indeed be expressed as the reductions
of covariant ten-dimensional ones.\footnote{It is worth emphasising
that this requirement of reinterpreting the T-duality transformed
type IIA string action as a {\it covariant} ten-dimensional action
allows one to {\it deduce} the existence of the type IIB string with
its chiral fermions and self-dual 5-form.}  To see how this works, let us
consider a sample calculation in detail.

    The dimensional reduction of $Q^\4_{mn}$ gives rise to a number of
terms, including the contribution to $Q_{m'0}$ of the 3-form $F_{m'n'p'}$
that comes from $F_{mnpq}$.  From (\ref{rrpq}), we see that this
contribution will be
\be
Q^\4_{m'0}=\ft{\im}{12} e^\phi\, \btheta^{({\sst1}}\, 
\Gamma_{m'}{}^{p' q' r'}\,
\, \theta^{\sst2 )} F_{p'q'r'0}+\cdots\,.
\ee
In terms of the type IIB spinors this can be seen to
be
\be
Q^\4_{m'0}=\ft{\im}{12} e^{\fft34 \phi-\fft{\sqrt7}{4}\varphi}\, 
\bttheta^{[{\sst1}}\, \Gamma_{m'}{}^{p'q'r'}\, \Gamma_0\, 
\ttheta^{\sst2 ]}\, F_{p'q'r'}+\cdots\,,
\ee
where the changed dilaton prefactor results from the conversion of
the $D=10$ tangent-space components of $F_{p'q'r'0}$ to the $D=9$ 
tangent-space components of $F_{p'q'r'}$.  From (\ref{bush2}),
$Q^\4_{m'0}$ is related to the $(m'0)$ components of an antisymmetric
term $\wtd P_{mn}$ in the T-duality transformed theory.  One can
easily see that it arises from the reduction of the ten-dimensional
quantity
\be
\wtd P^\3_{mn} = -\ft{\im}{12}\, e^{\fft34\phi-\fft{\sqrt7}{4}\varphi}\,  
\bttheta^{[{\sst 1}}\, \Gamma_{[m}\,\Gamma^{pqr}\, \Gamma_{n]}\, 
\ttheta^{{\sst 2}]}\, \wtd F_{pqr}\,,
\ee
where $\wtd F_{p'q'r'}=-F_{p'q'r'}$.  (The need for the minus sign
becomes apparent after following a complete chain of analogous
T-duality transformations.)  In fact if we look at the complete set of
T-duality transformations, we find that the scalars $\phi$ and
$\varphi$ occur in the same combination
$\ft34\phi-\ft{\sqrt7}{4}\varphi$ in all the $\wtd P_{mn}$ and $\wtd
Q_{mn}$ expressions.  Noting that in the T-duality transformed theory
there should be a single scalar $\td\phi$ in $D=10$, it is therefore
natural to define it to be
\be
\td\phi = \fft34\phi-\fft{\sqrt7}{4}\varphi\,.\label{dil1}
\ee
We now recall that in the Kaluza-Klein reduction (\ref{kkdecomp}) for
the metric of the T-transformed theory, the $\td g_{00}$ component was
parametrized in terms of $\td\lambda$, which was related to $\phi$
and $\varphi$ by $\td\lambda = -\ft14\phi -\ft{\sqrt7}{4}\varphi$ (see
(\ref{bush1})).  Since this is not orthogonal to $\td\phi$, it is
natural to reparametrize it in terms of $\td\phi$ and a second linear
combination $\td\varphi$ of $\phi$ and $\varphi$ that {\it is}
orthogonal to $\td\phi$, namely 
\be
\td\varphi = -\fft{\sqrt7}{4} \phi -\fft34 \varphi\,.\label{dil2}
\ee
In terms of these two orthogonal fields, we now have $\td\lambda =
\ft14 \td\phi +\ft{\sqrt7}{4}\td\varphi$.  This is identical in form
to the original untilded $\lambda$ in the Kaluza-Klein metric
decomposition (\ref{kkdecomp}) for the type IIA theory.
The relations (\ref{dil1}) and (\ref{dil2}) which we have derived here
are precisely the transformations that relate the dilatonic scalars in
the nine-dimensional reductions of the type IIA and type IIB
supergravities (see, for example, \cite{llps}).  

    It is interesting to note that our derivation of the relations
(\ref{dil1}) and (\ref{dil2}) in the present Green-Schwarz formalism
was a purely classical one.  By contrast, in a standard NSR sigma
model formulation of T-duality, the derivation of the dilaton
transformations requires a detailed consideration of how conformal
invariance can be preserved under quantisation
\cite{alvarez2lozano,gpr}.  The essential difference in the
Green-Schwarz formalism that has allowed us to obtain the dilaton
transformations from purely algebraic classical considerations is the
presence of R-R background field strengths in the string action.

    After carrying out the entire chain of T-duality transformations
(\ref{bush2}), involving repeated steps paralleling those that we have
illustrated above, we find that the R-R couplings in $D=10$ become
\bea
\wtd Q^\1_{mn} &=& -\ft{\im}{2} e^{\td\phi}\, 
\bttheta^{[{\sst 1}}\, \Gamma_{(m}\,
\Gamma^p\, \Gamma_{n)}\, \ttheta^{{\sst 2}]}\, \wtd F_p\,,\nn\\
\wtd P^\1_{mn} &=& \ft{\im}{2}\,e^{\td\phi}\,  
\bttheta^{({\sst 1}}\, \Gamma_{[m}\,
\Gamma^p\, \Gamma_{n]}\, \ttheta^{{\sst 2})}\, \wtd F_p\,,\nn\\
\wtd Q^\3_{mn} &=& \ft{\im}{12}\, e^{\td\phi}\,  
\bttheta^{({\sst 1}}\, \Gamma_{(m}\,
\Gamma^{pqr}\, \Gamma_{n)}\, 
\ttheta^{{\sst 2})}\, \wtd F_{pqr}\,,\label{rrtpq}\\
\wtd P^\3_{mn} &=& -\ft{\im}{12}\, e^{\td\phi}\,  
\bttheta^{[{\sst 1}}\, \Gamma_{[m}\,
\Gamma^{pqr}\, \Gamma_{n]}\, \ttheta^{{\sst 2}]}\, \wtd F_{pqr}\,,\nn\\
\wtd Q^\5_{mn} &=& -\ft{\im}{480}\, e^{\td\phi}\,  
\bttheta^{[{\sst 1}}\, \Gamma_{(m}\,
\Gamma^{p_1\cdots p_5}\, \Gamma_{n)}\, 
\ttheta^{{\sst 2}]}\, \wtd F_{p_1\cdots p_5}\,, \nn\\
\wtd P^\5_{mn} &=& \ft{\im}{480}\, e^{\td\phi}\,  
\bttheta^{({\sst 1}}\, \Gamma_{[m}\,
\Gamma^{p_1\cdots p_5}\, \Gamma_{n]}\, 
\ttheta^{{\sst 2})}\, \wtd F_{p_1\cdots p_5}\,.\nn
\eea
Note that the chirality of the $\ttheta^{\sst A}$ fermions implies
that the 5-form $\wtd F_\5$ is self-dual.  Note also that
$\wtd F_\1=d\chi$, where $\chi$ is the axionic scalar of the type IIB theory.

    We find that the type IIB 1-form, 3-form and 5-form field
strengths appearing in (\ref{rrtpq}) are related to the 2-form and
4-form field strengths of the type IIA action (\ref{type2alag}) as
follows.  Reducing the $D=10$ type IIA fields as
\be
F_\4 \longrightarrow F_\4 + F_\3\wedge (dY+\cA_\1)\,,\qquad
F_\2 \longrightarrow F_\2 + F_\1\wedge (dY+\cA_\1)\,,
\ee
and the $D=10$ type IIB fields as
\bea
&&\wtd F_\3 \longrightarrow \wtd F_\3 + \wtd F_\2\wedge 
(dZ+\wtd \cA_\1)\,,\qquad
\wtd F_\1 \longrightarrow \wtd F_\1\,,\nn\\
&&\wtd F_\5 \longrightarrow \wtd F_\5 + \wtd F_\4\wedge(dZ+\wtd
\cA_\1)\,,\label{2brrred}
\eea
where $\wtd F_\5$ and $\wtd F_\4$ in $D=9$ are related by Hodge
duality, we find that the nine-dimensional R-R fields must be identified
as follows:
\be
\wtd F_\4 = F_\4\,,\qquad \wtd F_\3 = - F_\3\,,\qquad
\wtd F_\2 = F_\2\,,\qquad \wtd F_\1 = -F_\1\,.
\ee
This particular pattern of identifications, including the alternating
signs, is precisely in agreement with the results for R-R T-duality
that one finds at the field-theory level.  (See, for example,
\cite{llps}.) 

    The discussion of the T-duality relations for the NS-NS
contributions at $O(\theta^2)$ proceeds in an analogous fashion.
Following a similar strategy to that described above, we find that 
the various background fields can be rewritten, after applying the
T-duality transformations (\ref{bush1}), in the $D=10$ covariant forms 
\bea
\wtd Q^{\ns}_{mn} &=& \ft{\im}4 
[\bttheta^{\sst1}\Gamma_{(m}{}^{pq}\, \ttheta^{\sst1} -
\bttheta^{\sst2}\Gamma_{(m}{}^{pq}\, \ttheta^{\sst2}
]\,
\wtd F^\ns_{n)pq}  +\ft{\im}2 [\bttheta^{\sst1} \Gamma_{(m}\,
\Gamma_{|pq|}\, \ttheta^{\sst1} + 
\bttheta^{\sst2} \Gamma_{(m}\,
\Gamma_{|pq|}\, \ttheta^{\sst2}
]\, \td\omega_{n)}{}^{pq}\,,\nn\\
\wtd P^{\ns}_{mn} &=& -\ft{\im}4 
[\bttheta^{\sst1}\Gamma_{(m}{}^{pq}\, \ttheta^{\sst1} +
\bttheta^{\sst2}\Gamma_{(m}{}^{pq}\, \ttheta^{\sst2}
]\,
\wtd F^\ns_{n)pq} -\ft{\im}2 [\bttheta^{\sst1} \Gamma_{(m}\,
\Gamma_{|pq|}\, \ttheta^{\sst1} - 
\bttheta^{\sst2} \Gamma_{(m}\,
\Gamma_{|pq|}\, \ttheta^{\sst2}
]\, \td\omega_{n)}{}^{pq}\,,\nn\\
\wtd {\cal G}_{i m} &=& - \im\, [\bttheta^{\sst1}\Gamma_m\, 
\del_i\ttheta^{\sst 1} +
\bttheta^{\sst2}\Gamma_m\, \del_i\ttheta^{\sst 2}
]\,,\label{nstpq}\\
\wtd {\cal B}_{im} &=& -
\im\, [\bttheta^{\sst1}\Gamma_m\, \del_i\ttheta^{\sst 1} -
\bttheta^{\sst2}\Gamma_m\, \del_i\ttheta^{\sst 2}
]\,,\nn
\eea
Reducing the ten-dimensional NS-NS fields to $D=9$ in the standard
Kaluza-Klein style, with $F^\ns_\3 \longrightarrow F^\ns_\3 +
F^\ns_\2\wedge dz$, we find that here, as in the previous
$O(\theta^0)$ discussion, the nine-dimensional type IIA and type IIB 
NS-NS fields are related in the standard way, with 
\be
\wtd F^\ns_\2 = \cF_\2\,,\qquad \wtd \cF_\2 = F^\ns_\2\,,
\ee
where $\cF_\2=d\cA_\1$ and $\wtd \cF_\2=d\cA_\1$.  This is just the
standard interchange of Kaluza-Klein and winding vectors.  

     A number of comments are now in order.  Firstly, we may observe
by comparing (\ref{nspq}) and (\ref{nstpq}) that the NS-NS terms at
$O(\theta^2)$ are identical in form in the type IIA and type IIB
Green-Schwarz actions.  This is in accordance with what one would
expect; it was already seen, of course, at $O(\theta^0)$.  A
particular consequence of this is that the $\del_i\ttheta^{\sst A}$
terms and the $\ft14 \del_i X^\mu\, \td\omega_\mu{}^{mn}\,
\Gamma_{mn}\ttheta^{\sst A}$ terms, which were separated into
contributions in $Q^\ns_{mn}$, $P^\ns_{mn}$, ${\cal G}_{im}$ and
${\cal B}_{im}$ during the implementation of the T-duality
transformation, have re-assembled themselves to make the spacetime
Lorentz-covariant derivative $D_i\, \ttheta^{\sst A}$ again.

     A second observation is that in the R-R sector, the expressions
for $Q_{mn}$ and $P_{mn}$ in the type IIA theory in (\ref{rrpq}) and
in the type IIB theory (\ref{rrtpq}) all have the same structural
form.  In all cases, the general structure for the coupling to a $p$-form
is
\be
Q^{\sst{(p)}}_{mn} \sim \fft{\im}{2\, p!}\, e^\phi\, \btheta^{\sst 1}\,
\Gamma_{(m}\, \Gamma^{q_1\cdots q_p}\, \Gamma_{n)}\, \theta^{\sst 2}\,
F_{q_1\cdots q_p}\,,\qquad
P^{\sst{(p)}}_{mn} \sim \fft{\im}{2\, p!}\, e^\phi\, \btheta^{\sst 1}\,
\Gamma_{[m}\, \Gamma^{q_1\cdots q_p}\, \Gamma_{n]}\, \theta^{\sst 2}\,
F_{q_1\cdots q_p}\,.
\ee
(In the case of the 5-form field strength in the type IIB theory, the
proper handling of the self-duality constraint implies, as usual, 
that this term should enter with $\fft12$ of the canonical coefficient
for a $p$-form.)

    To close this section, we may summarise our results for the
ten-dimensional type IIB Green-Schwarz action, which takes the 
form\footnote{We have corrected one typographical sign error that
arose in the version of this paper in Nucl. Phys. {\bf B573} (2000) 149
and v1. of hep-th/9907202, in the $\ep^{ij}$ term on the second line.
We have also implemented the modified definition where $\td\theta^2$ is
Majorana, as discussed in the Addendum section \ref{addsec}.} 
\bea
\wtd{\cal L}_2\!\!\!\! &=&\!\!\!\!
 -\ft12 \sqrt{-h}\, h^{ij}\, \del_i X^\mu\, \del_j\,
X^\nu\, \wtd g_{\mu\nu} + \ft12\ep^{ij}\, \del_i\, X^\mu\, \del_j\, X^\nu\,
\wtd A_{\mu\nu} \nn\\ 
\!\!\!\!&&\!\!\!\!- \im\, \sqrt{-h}\, h^{ij} \, 
(\bttheta^{\sst1}\, \Gamma_\mu\, D_j\ttheta^{\sst1} +
\bttheta^{\sst2}\, \Gamma_\mu\, D_j\ttheta^{\sst2})\, \del_i
X^\mu
-\im\, \ep^{ij}\, (\bttheta^{\sst1}\, \Gamma_\mu\, D_j\ttheta^{\sst1} -
\bttheta^{\sst2}\, \Gamma_\mu\, D_j\ttheta^{\sst2})\, \del_i
X^\mu\nn\\
\!\!\!\!&&\!\!\!\!
-\ft{\im}8 \sqrt{-h}\, h^{ij}\, \del_i X^\mu\, \del_j X^\nu\,
(\bttheta^{\sst1} \, \Gamma_\mu{}^{\rho\sigma}\, \ttheta^{\sst1}
-\bttheta^{\sst2} \, \Gamma_\mu{}^{\rho\sigma}\, \ttheta^{\sst2})\,
\wtd F^\ns_{\nu\rho\sigma}\\
\!\!\!\!&&\!\!\!\!
-\ft{\im}8\ep^{ij}\,  \del_i X^\mu\, \del_j X^\nu\,
(\bttheta^{\sst1} \, \Gamma_\mu{}^{\rho\sigma}\, \ttheta^{\sst1}
+\bttheta^{\sst2} \, \Gamma_\mu{}^{\rho\sigma}\, \ttheta^{\sst2})\,
\wtd F^\ns_{\nu\rho\sigma}\label{type2baction}
\ +\im\, e^{\td\phi}\, \del_i X^\mu\, \del_j X^\nu\,\times \nn\\
\!\!\!\!&&\!\!\!\!
\Big[
\sqrt{-h}\, h^{ij}\, \Big(\ft{1}{4}\, \bttheta^{[\sst 1}\, \Gamma_\mu\,
\Gamma^\rho \,\Gamma_\nu\, \ttheta^{\sst2]}\, \del_\rho\chi 
-\ft{1}{24}\,  \bttheta^{(\sst 1}\, \Gamma_\mu\,
\Gamma^{\rho\sigma\lambda}\, \Gamma_\nu\,  \ttheta^{\sst2)}\, 
\wtd F_{\rho\sigma\lambda}
+\ft{1}{960}\,  \bttheta^{[\sst 1}\, \Gamma_\mu\,
\Gamma^{\rho_1\cdots\rho_5}\, \Gamma_\nu\,  
\ttheta^{\sst2]}\, \wtd F_{\rho_1\cdots\rho_5}
\Big)\nn\\
\!\!\!\!&&\!\!\!\! 
+\ep^{ij}\, \Big(\ft{1}{4}\, \bttheta^{(\sst 1}\, \Gamma_\mu\,
\Gamma^\rho\, \Gamma_\nu\,  \ttheta^{\sst2)}\, \del_\rho\chi 
-\ft{1}{24}\,  \bttheta^{[\sst 1}\, \Gamma_\mu\,
\Gamma^{\rho\sigma\lambda}\, \Gamma_\nu\, 
\ttheta^{\sst2]}\, \wtd F_{\rho\sigma\lambda}
+\ft{1}{960}\,  \bttheta^{(\sst 1}\, \Gamma_\mu\,
\Gamma^{\rho_1\cdots\rho_5}\, \Gamma_\nu\,  
\ttheta^{\sst2)}\, \wtd F_{\rho_1\cdots\rho_5}
\Big)\Big]\,.\nn
\eea

\section{Green-Schwarz Action for the Massive Type IIA Theory}

    It has been shown at the level of the low-energy effective
supergravities that the massive type IIA theory and the standard type
IIB theory are related {\it via} a T-duality transformation that
differs from the usual one only in that a generalised Scherk-Schwarz
reduction ansatz is now introduced for the axion $\chi$ of the IIB
theory \cite{bdgpt}.  In other words, one makes the more general
Kaluza-Klein ansatz
\be
\chi(x,z)\longrightarrow \chi(x) + m\, z\label{chired}
\ee
when reducing from $D=10$ to $D=9$, while keeping all other ans\"atze
unchanged.  Despite the linear dependence on $z$ in (\ref{chired}),
the usual requirement of $z$-independence of the higher-dimensional
action and field strengths is still satisfied, in view of the global
shift symmetry under $\chi\longrightarrow \chi\ +\ $constant.

     It is straightforward to implement the analogue of this generalised
reduction in the T-duality transformation of the previous section.
Thus we now replace the reduction procedure for $\wtd F_\1$ given in
(\ref{2brrred}) by
\be
\wtd F_\1 \longrightarrow \wtd F_\1 + m\, (dZ +\wtd \cA_\1)\,.
\ee
(As a consequence, $\wtd F_\1$ in $D=9$ is now given by $\wtd F_\1 =
d\chi -m\, \wtd\cA_\1$.)   We can now use the T-duality rules
(\ref{bush2}) in the direction opposite to that which we followed
previously, to deduce the form of the new terms $Q^\0_{mn}$ and
$P^\0_{mn}$ that we shall acquire in the type IIA string action.  As
usual, we find that after applying the T-duality transformation, the
nine-dimensional expressions can indeed be lifted to covariant
ten-dimensional ones.  By this means, we obtain the following new
terms in the ten-dimensional type IIA Green-Schwarz action:
\be
Q^\0_{mn} = \ft{\im}2 m\, e^\phi\, (\btheta^{\sst1}\, \Gamma_{(m}\,
\Gamma_{n)}\, \theta^{\sst1} + \btheta^{\sst2}\, \Gamma_{(m}\,
\Gamma_{n)}\, \theta^{\sst2})\,,\qquad
P^\0_{mn} = \ft{\im}2 m\, e^\phi\, (\btheta^{\sst1}\, \Gamma_{mn}\,
 \theta^{\sst1} - \btheta^{\sst2}\, \Gamma_{mn}\,
\, \theta^{\sst2})\,.
\ee
(In the expression for $Q^\0_{mn}$ we could, of course, replace
$\Gamma_{(m}\, \Gamma_{n)}$ by $\eta_{mn}$.)  Thus up to
$(O(\theta^2)$, we find that the Green-Schwarz action for the massive
type IIA string is given by
\be
{\cal L}_2^{\rm massive} = {\cal L}_2 - \ft{\im}4 m\,\del_i X^\mu\, \del_j
X^\nu\, e^\phi\,
\btheta\, (\sqrt{-h}\, h^{ij} -\ep^{ij}\, \Gamma_{11})\, \Gamma_\mu\,
\Gamma_\nu\, \theta\,,\label{type2amassive}
\ee
where ${\cal L}_2$ is given in (\ref{type2alag}), but with the field
strengths now given by
\bea
F_\2 &=& dA_\1 + m\, A_\2\ ,\qquad F_\3 =
dA_\2\ ,\nn\\
F_\4 &=& dA_\3 + A_\1\wedge dA_\2
+ \ft12 m\, A_\2\wedge A_\2
\,.\label{romfields0}
\eea

     It is often more appropriate to use a formalism where the mass term
of the IIA theory, which can be thought of as a 0-form field strength, 
is replaced by its dual 10-form field strength.  In this case the action
will be given instead by
\be
{\cal L}_2^{\rm massive} = {\cal L}_2 -\ft{\im}{4\times 10!}\, 
\del_i X^\mu\, \del_j
X^\nu\, e^\phi\,
\btheta\, (\sqrt{-h}\, h^{ij} -\ep^{ij}\, \Gamma_{11})\, \Gamma_\mu\,
\Gamma^{\rho_1\cdots\rho_{10}}\, \Gamma_\nu\, 
\theta\, F_{\rho_1\cdots \rho_{10}}\,.
\ee

\section{Massless/massive type IIA T-duality}

     In the previous section, we derived the T-duality transformation
between the type IIA and type IIB Green-Schwarz actions.  In
particular, if the 0-form $m$ vanishes, this relates the massless type
IIA and type IIB theories.  When $m$ is non-vanishing, the T-duality
maps from type IIB to the massive type IIA string theory.  This
T-duality has been discussed in Refs \cite{bdgpt,clpst,ortin2,llps} at
the level of the low-energy effective action, by considering the
Scherk-Schwarz reduction of type IIB supergravity.  This leads to the
expectation that there should be a T-duality that directly relates the
massless and massive type IIA string theories.  This issue was
explored in \cite{hull}, by considering the T-duality between the
D8-brane and the D6-brane in eight dimensions.  Since the D6-brane in
$D=8$ has, from an eleven-dimensional point of view, the internal
structure of either a circle bundle over a 2-torus \cite{llp} or a
2-torus bundle over a circle \cite{hull}, this would provide a
geometrical relation between M-theory and the massive type IIA string.

    Here, we shall consider the T-duality relations between the
massive and massless type IIA strings in detail, making use of the
results of the previous sections.  In particular, we have seen that
the results of T-duality transformations on the string variables
$(X^\mu,\theta)$ are implemented on the background fields by means of
standard nonlinear supergravity global symmetry transformations.
This would then also imply that a T-duality map between an original
massless string theory (with an appropriate non-trivial R-R background)
and the corresponding image massive theory will similarly be
implemented by a background-field transformation that can be read off
from previous work on effective field theories. In particular, for the
purposes of the present discussion we shall consider string theories
on backgrounds with two isometries, which are related to previous work
on $D=8$ effective supergravities.

     Accordingly, let us consider massless type IIA string theory on a
background with two $U(1)$ isometries, namely where all
spacetime fields are independent of two spacetime coordinates
$Y^a$, $a=1, 2$. Defining $V_i^a=\del_i Y^a$, we introduce the
Lagrange multiplier term
\be
{\cal L} = {\cal L}_0 + \epsilon^{ij} \del_i Z_a V_j^a
\,.\label{d8lag}
\ee
Next, we integrate out the two $V_i^a$.  Two possible cases arise,
according to whether or not the 2-form field strength $F_\2$ with
indices projected into the compactified directions vanishes. These two
cases correspond to massless or massive string backgrounds
respectively.

     Consider first the massless case. After performing a T-duality
transformation by integrating out the two $V_i^a$, the Lagrangian can once
again be put into the form of the type IIA massless string action, but
with background fields now subjected to a transformation which can be
identified as one of the perturbative symmetries of the $D=8$
effective supergravity theory.  After a final relabelling of the
Lagrange multipliers $(Z_1=\wtd Y^2,\,Z_2=-\wtd Y^1)$, this ``double''
T-duality transformation corresponds to an inversion and interchange
of the two toroidal radii; $R_1\longrightarrow 1/R_2$ and
$R_2\longrightarrow 1/R_1$.  Unlike the situation where one performs a
T-duality inversion on just a single circle, which maps the type IIA
theory into the IIB theory,\footnote{Note that in the case of bosonic
or type I strings, this single-inversion transformation {\it is} a
symmetry, and it enlarges the total T-duality group from $SO(2,2)$ to
$O(2,2)$.  See \cite{new1} for a recent discussion, at the level of
supergravity action, of how R-R potentials transform under the $SO(d,d)$
group.} this double T-duality inversion maps the type IIA theory back
into itself.  In other words, the transformed theory can still be
lifted up to the type IIA massless string action in $D=10$, so this
constitutes a genuine element of the symmetry group of the type IIA
string.

     As with the single-inversion T-duality discussed in section 3,
the effect of the double-inversion T-duality on the background fields
is identical to a standard transformation of the corresponding
lower-dimensional effective field theory.  In the single-inversion
case, this was the well-known redefinition (not a symmetry) that
relates the fields of the type IIA and type IIB forms of
nine-dimensional supergravity.  In the double-inversion case, the
corresponding field-theory transformation {\it is} a symmetry of the
eight-dimensional supergravity, which is contained in the perturbative
T-duality $SO(2,2)$ subgroup of the $SL(3,\R)\times SL(2,\R)$
Cremmer-Julia symmetry of maximal eight-dimensional
supergravity.\footnote{Since we shall be discussing symmetry groups
both in their continuous forms at the level of the effective field
theories, and in their discretised forms in the string theories, we
shall tend to refer to them always in their continuous forms, with
their discretisations to integer coefficients in the string-theory
context being understood.}

     Before we identify which $D=8$ supergravity symmetry
transformation the T-duality transformation produces, let us recall
the structure of the T-duality symmetry group. The perturbative
T-duality group that arises for spacetimes with the isometries of a
$d$-torus is $SO(d,d)$. This is a subgroup of the Cremmer-Julia
supergravity symmetry group for the effective field theory obtained by
dimensionally reducing from $D=10$ on such a torus; in the case of
$D=8$ maximal supergravity, the Cremmer-Julia symmetry group is
$SL(3,\R)\times SL(2,\R)$. The corresponding $SO(2,2)$ perturbative
symmetry is isomorphic to $SL(2,\R)_1\times SL(2,\R)_2$, where the
$SL(2,\R)_2$ factor is the second factor in the Cremmer-Julia group
and the $SL(2,\R)_1$ factor is a subgroup of the Cremmer-Julia
$SL(3,\R)$. $SL(2,\R)_1$ has its origin in general coordinate
transformations of the 2-torus. In fact, the full residuum of the
``internal'' general coordinate transformations is $GL(2,\R)\sim
\R\times SL(2,\R)$, but the $\R$ factor becomes part of
$SL(2,\R)_2$. The remaining two $SL(2,\R)_2$ generators arise from an
$O(2)$ transformation independent of the general-coordinate $GL(2,\R)$
and from constant shifts of the axion $\chi=A_\0$ that comes from the
internal components of the NS-NS 2-form in $D=10$. (See, for example,
\cite{giveonrocek} for a discussion of $O(d,d)$ T-duality for string
actions on nontrivial NSR backgrounds.)

    Now we shall identify which $SO(2,2)$ transformation has been
generated by the double-inversion T-duality. Specifically, we shall
show that it is the element of the $SL(2,\R)_2$ factor represented by
the matrix
\be
\Lambda=\pmatrix{0 & 1\cr -1 & 0}\,.\label{z2el}
\ee

    To see how this works, let us consider the double-inversion
T-duality transformation in detail.  We already saw in section 3 that
the usual presentation of the single-inversion T-duality
transformation leads to rather opaque transformation rules of the form
(\ref{bosbus}), which assume the more elegant form (\ref{bush1}) when
the ten-dimensional fields are decomposed in terms of nine-dimensional
fields in a geometrically-natural way.  Specifically, it is the
standard Kaluza-Klein ansatz that provides this natural framework for
this decomposition.  This advantage becomes all the more persuasive in
the present context, where we wish to implement the double-inversion
T-duality transformation corresponding to integrating out both of the
auxiliary fields $V_i^a$ in (\ref{d8lag}).  Thus we begin by writing
the ten-dimensional string metric in the Kaluza-Klein decomposition
appropriate to the assumed form where there are two $U(1)$ isometries
on a 2-torus:
\be
d\hat s_{10}^2 = ds_8^2 + 
e^{-f+\psi}\, (dY^1 + \cA_\1^1 + \cA^1_{\0 2}\,
dY^2)^2 + e^{-f-\psi}\, (dY^2 + \cA_\1^2)^2\,,\label{d8red2}
\ee
where all fields are independent of the two toroidal coordinates
$(Y^1,Y^2)$, and we have defined
\be
f\equiv -\ft12 \phi -\ft3{2\sqrt7}\varphi_1 -\sqrt{\ft37} \varphi_2\,,
\qquad \psi\equiv \ft2{\sqrt7} \varphi_1 -\sqrt{\ft37}\varphi_2\,.
\label{fpsidef}
\ee
Here, $\varphi_1$ and $\varphi_2$ are the usual dilatonic scalars
coming from the reduction steps to $D=9$ and $D=8$ respectively.  As
we shall see, the combination $f$ will be the canonically-normalised
dilaton in the $SL(2,\R)/O(2)$ scalar coset describing the
$SL(2,\R)_2$ global symmetry.  The combination $\psi$, also
canonically normalised, is orthogonal to $f$ and lives in the
$SL(3,\R)/O(3)$ part of the total scalar manifold.  We also make a
standard Kaluza-Klein decomposition for the 2-form potential
$A_{\mu\nu}$:
\be
\hat A_\2 = A_\2 + A_{\1 a}\wedge dY^a + \chi\, dY^1\wedge dY^2\,.
\ee
In fact $\chi$ will turn out to be the axion in the $SL(2,\R)_2/O(2)$
scalar coset.

    We are now ready to implement the double-inversion T-duality
transformation.  Thus we begin with the sigma-model Lagrangian
\be
{\cal L}_0 = -\ft12 \sqrt{-h}\, h^{ij}\, \del_i X^\mu\, \del_j X^\nu\,
G_{\mu\nu} + \ft12 \ep^{ij} \, \del_i X^\mu\, \del_j X^\nu\,
A_{\mu\nu}\,,\label{d8lag0}
\ee
and make the assumption that all fields are independent of $Y^a=(Y^1,
Y^2)$, where we decompose the coordinates as $X^\mu = (X^{\mu'}, Y^a)$.
Then replacing $\del_i Y^a$ by $V_i^a$, adding the Lagrange multiplier
term as in (\ref{d8lag}), and integrating out the two $V_i^a$, we arrive
at a T-duality transformed Lagrangian which again has the same general
form as (\ref{d8lag0}), with $(Z_1,Z_2)$ now playing the r\^oles of the
two toroidal coordinates.  In fact, it is natural to relabel these in
terms of two tilded coordinates $\wtd Y^a$, according to the rule $Z_1
= \wtd Y^2$, $Z_2 = - \wtd Y^1$.  As we indicated above, at the
intermediate stages of the calculation the results
are cumbersome.
 However, they become rather simple when expressed in terms
of the fields appearing in the Kaluza-Klein decomposition, and we shall
present only these final results here.  To do so, it is useful first
to make the following redefinitions:
\bea
\cA'{}_\1^1 &=& \cA_\1^1 -\, \cA^1_{\0 2}\, \cA_\1^2\,,\qquad
A_{\1 1}' = A_{\1 1} + \chi\, \cA_\1^2\,,\nn\\
A_{\1 2}' &=& A_{\1 2} - \chi\, {\cA_\1^1}'\,,\qquad
A_{\2}' = A_\2 -\chi\, {\cA_\1^1}'\wedge \cA_\1^2\,.
\eea

    After the double-inverse T-duality transformation, we find that
the sigma-model Lagrangian can be recast in the form (\ref{d8lag0}),
using tilded fields which are related to the original untilded ones as
follows.  First of all, in the scalar sector we find
\bea
e^{-\td f} &=& \fft{ e^{-f}}{e^{-2f} + \chi^2}\,,\qquad 
\wtd\chi = -\fft{\chi}{e^{-2f} + \chi^2}\,, \label{scalars}\\
\wtd\psi &=& \psi\,,\qquad \wtd\cA^1_{\0 2} = \cA^1_{\0 2}\,.\nn
\eea
This shows that the dilaton $f$ and the axion $\chi$ have transformed
under precisely the $SL(2,\R)$ matrix $\Lambda$ given in
(\ref{z2el}), where $\Lambda=\sst{\pmatrix{a&b\cr c&d}}$ acts by
fractional linear transformations $\td\tau=(a \tau+b)/(c \tau+d)$ on
$\tau\equiv \chi + \im\, e^{-f}$.  On the other hand the other
dilatonic scalar combination $\psi$, and the axion $\cA^1_{\0 2}$
associated with $SL(2,\R)_1$, are inert. We find that the remaining fields
are transformed in the following way:
\bea
\wtd\cA'{}_\1^1 &=& - A_{\1 2}'\,,\qquad 
\wtd\cA_\1^2 = A_{\1 1}'\,,\nn\\
\wtd A_{\1 1}' &=& - \cA_\1^2\,,\qquad 
\wtd A_{\1 2}' = {\cA_\1^1}'\,,\label{forms}\\
\wtd A_\2' &=& A_\2' + {\cA_\1^1}'\wedge A_{\1 1}' +
 {\cA_\1^2}\wedge A_{\1 2}'\,,\nn\\
\td g_{\mu'\nu'} &=& g_{\mu'\nu'}\,.\nn
\eea
Note that here $g_{\mu'\nu'}$ denotes the eight-dimensional
string-frame metric $ds_8^2$ appearing in the Kaluza-Klein
decomposition (\ref{d8red2}).  It is therefore related to $G_{\mu'\nu'}$
by $g_{\mu'\nu'} = G_{\mu'\nu'} + e^{-f}\, (e^\psi\, \cA^1_{\mu'}\,
\cA^1_{\nu'} + e^{-\psi}\, \cA^2_{\mu'}\, \cA^2_{\nu'})$.  

    The transformations (\ref{scalars}) and (\ref{forms}) describe a
symmetry of the truncation of the eight-dimensional supergravity to
its NS-NS sector.  In fact, when appropriately augmented by
transformations for the R-R fields, it is a symmetry of the full
theory (see the appendix for a complete discussion); as we discussed
above, it is the $Z_2$ subgroup of the $SL(2,\R)_2$ factor of the
T-duality symmetry given in (\ref{z2el}).

     So far, we have demonstrated the massless/massive type IIA
T-duality at the level of the NS-NS background fields, in what is a
rather natural generalisation of the standard discussion for IIA/IIB
duality.  Since we are working in the Green-Schwarz formalism, this
string-theoretic derivation of massless/massive IIA T-duality can be
extended to the R-R sector too, as we did for IIA/IIB T-duality in
section 3.  To do this, we again start from a generic Lagrangian of
the form (\ref{genlag}), with $G_{\mu\nu}$ and $B_{\mu\nu}$ containing
$O(\theta^2)$ terms $Q_{\mu\nu}$ and $P_{\mu\nu}$ as in (\ref{pqdef}),
together also with the terms ${\cal G}_{j\mu}$ and
${\cal B}_{j\mu}$ containing the $\btheta\del_j\theta$ 
at $O(\theta^2)$.  Following the lessons
learned in section 3.1, where we saw that the relations between the
lower-dimensional components of the original and the T-duality
transformed $O(\theta^2)$ tensors are much
simpler if written in terms of tangent-space indices, we follow the
same strategy here.  After algebra of considerable complexity, we
arrive at the following expressions for the tilded $O(\theta^2)$
backgrounds in terms of the original ones:
\bea
&&\wtd Q_{m'n'} = Q_{m'n'}\,,\qquad \wtd P_{m'n'} = P_{m'n'}
\,, \qquad \wtd {\cal G}_{im'} = {\cal G}_{im'}\,, \qquad
\wtd{\cal B}_{im'} = {\cal B}_{im'}\,,\nn\\
&&\wtd Q_{m' a} = e^{-\fft12(\td f+f)}\, \Big( -\ep_{ab}\, P_{m' b} +
\chi\, e^f\, Q_{m' a} \Big)\,,\nn\\
&&\wtd P_{m' a} = e^{-\fft12(\td f+f)}\, \Big( -\ep_{ab}\, Q_{m' b} +
\chi\, e^f\, P_{m' a} \Big)\,,\label{d8rrtduality}\\
&&\wtd Q_{12} =Q_{12}\,,\qquad \wtd Q_{11} -\wtd Q_{22} = Q_{11}
-Q_{22}\,,\nn\\
&&\wtd Q + \wtd\chi\, e^{\td f}\, \wtd P_{12} = 
-Q - \chi\, e^f\, P_{12}\,,\qquad
\wtd P_{12} - \wtd\chi\, e^{\td f}\, \wtd Q = -P_{12} + \chi\,
e^f\, Q\,,\nn\\
&&\wtd{\cal G}_{i a} =e^{-\fft12(\td f+f)}\, \Big(\ep_{ab}\, {\cal
B}_{i b} + \chi\, e^f\, {\cal G}_{i a}\Big)\,,\qquad
\wtd{\cal B}_{i a} =e^{-\fft12(\td f+f)}\, \Big(\ep_{ab}\, {\cal
G}_{i b} + \chi\, e^f\, {\cal B}_{i a}\Big)\,,\nn
\eea
Our notation here is that $m'$ denotes a tangent-space index
restricted to the
eight-dimensional subspace, and $a$ and $b$ denote tangent-space
indices ranging over the two values corresponding to the directions of
the two $U(1)$ isometries.  The explicit numerical indices 1 and 2
similarly denote these two tangent-space index values.  The quantity
$Q$ is defined by $Q\equiv \ft12 (Q_{11} + Q_{22})$, and likewise
$\wtd Q\equiv \ft12(\wtd Q_{11} + \wtd Q_{22})$.  

     In the appendix, we derive the complete $SL(2,\R)$
transformations of the NS-NS and R-R fields of the eight-dimensional
supergravity.  In the NS-NS sector, the $Z_2$ subgroup corresponding
to (\ref{z2el}) coincides precisely with the T-duality relations that
we derived in (\ref{scalars}) and (\ref{forms}) by string-theoretic
methods.  We have verified that this is also true in the R-R sector,
namely that $Z_2$ transformations of the R-R fields agree with the
transformations that follow from the $O(\theta^2)$ relations
(\ref{d8rrtduality}).  To demonstrate this, one has to follow steps
analogous to those that we presented for the case of type IIA/IIB
T-duality in section 3.2.  Note that the fermionic coordinates
$\theta$ also undergo a transformation under the double T-duality,
with $\ttheta= e^{-\fft12\a\, \Gamma_{11}\, \Gamma_1\, \Gamma_2}\,
\theta$, where $\sin\a=e^{-\ft12\td f-\ft12 f}$.  This corresponds to
a compensating transformation in the $O(2)$ denominator subgroup of the
$SL(2,\R)_2/O(2)$ coset.

        Now let us consider the case where $F_{\0 12}=m$ is allowed to
be a non-vanishing part of the background for the massless type IIA
theory.  (This corresponds, at the level of the effective field
theory, to a Scherk-Schwarz reduction of the axion $A_{\0 1}$ in $D=9$
that comes from the R-R 1-form $A_\1$ in $D=10$.)  After integrating
over the two $V^a_i$ auxiliary fields in (\ref{d8lag}), the resulting
Lagrangian no longer allows a direct interpretation as a dimensional
reduction of the massless ten-dimensional type IIA string.  By this we
mean that, unlike in the $F_{\0 12}=0$ discussion above, the tilded
backgrounds of the T-duality transformed theory cannot be directly
interpreted as the fields appearing in the given Kaluza-Klein
dimensional-reduction ansatz.  For example, the tilded 1-forms
$\wtd\cA_\1^a$ and $\wtd A_{\1 a}$, and the tilded dilatonic scalars
$\vec{\td\phi}$ , arising after integrating over the two $V_i^a$,
which in the $F_{\0 12}=0$ case could respectively be interpreted as
the Kaluza-Klein vectors, winding vectors and dilatons in the
Kaluza-Klein reduction ansatz for the type IIA theory, cannot be so
interpreted once $F_{\0 12}=m\ne0$.

    The reason why this has happened is that the $SL(2,\R)_2$ symmetry
is broken by the cosmological potential that arises in the
Scherk-Schwarz reduction with $F_{\0 12}=m\ne0$ \cite{clpst}.  
The full Lagrangian is given in the appendix; here, we shall just
consider the relevant terms in order to illustrate the point.  The
cosmological term, together with the kinetic terms for the scalars of
the $SL(2,\R)_2/O(2)$ coset, is given by
\be
{\cal L} = -\ft12 e\, m^2 \, e^{f+\fft{4}{\sqrt3}\sigma} 
-\ft12 e\, (\del f)^2
-\ft12 e\, e^{2f}\, (\del \chi)^2\,,\label{cosmo1}
\ee
where 
\be
\sigma\equiv \ft{\sqrt3}{2} \phi -\ft{\sqrt3}{2\sqrt7}\varphi_1
-\ft1{\sqrt{7}}\varphi_2\label{sigdef}
\ee
is the third linear combination of the three dilatonic scalars; it
is canonically normalised and is orthogonal to $f$ and to $\psi$
defined in (\ref{fpsidef}).  We can think of the cosmological term as
being the ``kinetic term'' for the 0-form field strength $F_{\0
12}=m$.  The key point that distinguishes this from the case when
$m=0$ is that whereas all the other field strengths are either
invariant under $SL(2,\R)_2$ or else they occur in doublet pairs, here
we have a single term which is not invariant under $SL(2,\R)_2$, thus
breaking the symmetry.  Nonetheless, we can still view $SL(2,\R)_2$ as
defining a set of field redefinitions, albeit ones that now change the
form of the Lagrangian.  In particular, it is easy to see that for a
general $SL(2,\R)_2$ transformation of the backgrounds, for which
\be
e^{-f}\longrightarrow \fft{e^{-f}}{(c\chi+d)^2 +c^2\, e^{-2f}}\,,\qquad
\chi \longrightarrow \fft{(a\chi+b)(c\chi+d)+ a\, c\, 
     e^{-2f}}{(c\chi+d)^2 + c^2\, e^{-2f}}\,,
\ee
the Lagrangian (\ref{cosmo1}) transforms into
\be
{\cal L} = -\ft12 e\, m_1^2 \, e^{-f+\fft{4}{\sqrt3} \sigma}
-\ft12 e\, (m_2 + m_1\, \chi)^2\,e^{f+\fft{4}{\sqrt3} \sigma} 
 -\ft12 e\, (\del f)^2
-\ft12 e\, e^{2f}\, (\del \chi)^2\,,\label{cosmo2}
\ee
where
\be
\pmatrix{m_1\cr m_2} = \Lambda^T\, \pmatrix{0\cr m}\,,
\ee
with $\Lambda= \sst{\pmatrix{a&b\cr c&d}}$. 

     Our specific $SL(2,\R)_2$ transformation that results from the
double-inversion T-duality has $\Lambda$ given by (\ref{z2el}), and
hence the Lagrangian (\ref{cosmo2}) becomes
\be
{\cal L} = -\ft12 e\, m^2 \, e^{-f+\fft{4}{\sqrt3}\sigma}
-\ft12 e\, m^2\, \chi^2\,e^{f+\fft{4}{\sqrt3}\sigma} 
 -\ft12 e\, (\del f)^2
-\ft12 e\, e^{2f}\, (\del \chi)^2\,,\label{cosmo3}
\ee
It is clear that this is different from the original form of the
Lagrangian (\ref{cosmo1}), showing that the mass term breaks not only
the general $SL(2,\R)_2$ symmetry but also the specific $Z_2$ subgroup
corresponding to the double-inversion T-duality (\ref{z2el}).  
However, the T-duality transformed  backgrounds {\it
do} still allow themselves to be directly lifted to a covariant theory
in $D=10$, but now it is to the massive IIA theory, rather than to the
massless theory.  In other words we {\it
can} directly interpret $\wtd\cA_\1^a$ as the Kaluza-Klein vectors,
$\wtd A_{\1 a}$ as the winding vectors, and $\vec{\td\phi}$ as the
dilatons coming from the reduction ansatz for the massive IIA theory.
In particular, we can see this from the form of the cosmological term
in (\ref{cosmo3}).  From the definitions of $f$ and $\sigma$, we see
that the cosmological term is  
\be
-\ft12 m^2\, e^{\fft52\phi -\fft1{2\sqrt7}\varphi_1 -\fft1{\sqrt{21}}
\varphi_2}\,, 
\ee
which is precisely what one obtains by performing a Kaluza-Klein
reduction of the cosmological term $-\ft12 m^2\, e^{\fft52\phi}$ in
the massive IIA theory.  The mass term for the axion $\chi$ in
(\ref{cosmo3}) can also be understood from this viewpoint, since it is
nothing but the Kaluza-Klein reduction of the mass term for the
2-form $A_\2$ in the ten-dimensional massive IIA theory.

\section{Conclusions}

    In this paper we studied the T-duality maps involving the type IIA
and type IIB string theories at the level of the Green-Schwarz
sigma-model string actions.  This approach provides a
string-theoretical derivation of T-dualities, and sheds light on the
subsequent web of dualities between the massless and massive type IIA
and type IIB string theories.
  
     As a prerequisite for these analyses, we needed explicit
(component) forms of the Green-Schwarz actions, in backgrounds
including R-R as well as NS-NS massless fields.  While the superfield
form of the type IIA action in curved backgrounds has been given in
\cite{howe}, the component form, needed for the application of
T-duality transformations, was unknown. The component form of the
analogous type IIB Green-Schwarz action was also previously unknown.
In our derivation of the type IIA Green-Schwarz action, we employed a
double dimensional reduction of the eleven-dimensional supermembrane
action, and making use of explicit results for the supervielbein and
3-form superpotential, derived up to order $\theta^2$ in the fermionic
fields \cite{dwpp}.  This enabled us to find an explicit and complete
form of the type IIA Green-Schwarz action with massless NS-NS {\it
and} R-R background fields, exact up to order $\theta^2$.
 
   This action provided our starting point for studying T-duality
transformations at the level of the Green-Schwarz action. We
generalised Buscher's T-duality procedure (implemented in
\cite{buscher} in an NSR formalism with only NS-NS background fields)
to the Green-Schwarz action, now involving spinor coordinates $\theta$
as well as R-R background fields.  Furthermore, we formulated the
T-duality transformation rules in terms of adapted background-field
parametrizations, thus obtaining more compact and elegant expressions
for the T-duality maps which have a natural geometrical
interpretation.

   These generalised T-duality transformations enabled us in turn to
derive the Green-Schwarz action for the type IIB string, giving it for
the first time with NS-NS {\it and} R-R background fields, exact up to
$O(\theta^2)$.
 
    Next, we constructed the string-theoretical T-duality map between
the type IIB and massive type IIA strings.  Starting with the type IIB
Green-Schwarz action, in a background with a $U(1)$ isometry in which
the 1-form field strength $F_\1=d\chi$ is allowed to take a
non-vanishing constant value in the direction of the isometry, and
applying the generalised T-duality transformation, we derived the
Green-Schwarz action for the massive IIA string, in arbitrary R-R and
NS-NS maslles backgrounds, again to order $\theta^2$. This provides a
string theoretical derivation of the T-duality transformation which,
at the level of the effective supergravities, corresponds to the
equivalence of the Scherk-Schwarz reduced type IIB and massive type 
IIA theories.

   In view of the fact that the type IIB string is itself related by
T-duality to the massless type IIA string, one can obtain a direct
T-duality relation between the massless and massive type IIA strings,
by considering them on backgrounds with the two $U(1)$ isometries of a
2-torus.  This was discussed by considering D8-branes and D6-branes in
the effective field theories in \cite{hull}; here, we were able to
give a string-theoretic description at the level of T-duality in the
Green-Schwarz actions.  In addition, we gave a construction of the
eight-dimensional $SL(3,\R)\times SL(2,\R)$ invariant supergravity
effective action that includes the two mass parameters $m_{1,2}$,
forming a doublet under the $SL(2,\R)$ factor. This can be interpreted
as a Scherk-Schwarz reduction of the type IIB string on two circles of
radii $R_1$ and $R_2$ (introducing mass parameters $m_1$ and $m_2$
respectively).  The T-duality transformation in turn relates this to
the massive type IIA string with one radius compactified on a regular
circle and another one Scherk-Schwarz reduced.

   Finally, we remark that our discussion of the T-duality mapping
between the massless and massive type IIA strings extends
straightforwardly, in a manner analogous to that discussed in
\cite{hull}, to a unified picture in which the massive type IIA string
emerges in an appropriate limit from the eleven-dimensional
supermembrane propagating in a background with the three $U(1)$
isometries of a 3-torus, which can be viewed either as an $S^1$
bundle over $T^2$ \cite{lav} or as a $T^2$ bundle over $S^1$ \cite{hull}.

\section{Addendum}\label{addsec}

   Since writing this paper, and its publication in Nucl. Phys. {\bf
B573} (2002) 149 and as v1. of hep-th/9907202, we have discovered a
small number of minor typographical errors.  We have also found that
there were various infelicitous choices of convention and notation
which, although in no way incorrect, did not contribute positively to
the elegance of the results.  We are therefore taking the opportunity,
while correcting the minor typographical errors, of improving some
notation and conventions.  We shall set out below what these changes
are.  For clarity, we shall refer to the versions of this paper
published in Nucl. Phys. {\bf B573} (2002) 149 and in v1. of
hep-th/9907202 as the ``old versions,'' and this current version will
be called the ``new version.''

   There were two typographical errors in the old versions that
affected our final expressions for the type IIA and type IIB
Lagrangians respectively.  In the type IIA Lagrangian
(\ref{type2alag}), a $\Gamma_{11}$ matrix was accidentally omitted
from the $F_{\nu\rho\sigma}$ terms.  In the type IIB Lagrangian
(\ref{type2baction}), the term involving $\im\, \ep^{ij}\,
(\bttheta^{\sst1}\, \Gamma_\mu\, D_j\ttheta^{\sst1} -
\bttheta^{\sst2}\, \Gamma_\mu\, D_j\ttheta^{\sst2})$ was accidentally
written as $\ep^{ij}\, (\bttheta^{\sst1}\, \Gamma_\mu\,
D_j\ttheta^{\sst1} + \bttheta^{\sst2}\, \Gamma_\mu\,
D_j\ttheta^{\sst2})$.  Both errors are corrected in this new version.

   Turning now to the question of conventions, we found that there were
two convention choices in the old versions of this paper that were not
helpful to the interpretation of our results. 

   Firstly, our starting point of the supermembrane action in
\cite{dwpp} is written using a convention in which the conjugation of
spinors involves the introduction of a factor of $\im$; \ie the
conjugate of a spinor $\psi$ is written as $\bar\psi=\im\,
\psi^\dagger\, \Gamma^t$, where $\Gamma^t$ denotes the Dirac matrix in
the time direction.  This has the consequence that, for example, the
Hermitean Dirac action is written as $\bar\psi\, \Gamma^\mu\, D_\mu\,
\psi$ rather than the more familiar $\im\, \bar\psi\, \Gamma^\mu\,
D_\mu\, \psi$.  To readjust to conventions that are more familiar, we
have therefore implemented the replacement of every occurrence of a
conjugate spinor by $\im$ times the conjugate spinor in all formula
in the old versions of this paper.  Thus we have replaced
\be
\bar\theta_{\rm old\ version} \longrightarrow \im\, 
\bar\theta_{\rm new\ version}\,,\label{hollandaisedressing}
\ee
{\it et cetera}.  

  A second inconvenience in our conventions was that when defining the
two chiral spinors $(\td\theta^1, \td\theta^2)$ of the type IIB theory
in terms of the spinor $\theta$ of type IIA that has chiral and
antichiral projections $(\theta^1, \theta^2)$, we defined
$\td\theta^2$ as $-\im\, \Gamma_0\, \theta^2$, rather than as $
\Gamma_0\, \theta^2$ as we are now doing in the this new version of
the paper, in equation (\ref{2ato2b}). (Recall that $\Gamma_0$ is the
Dirac matrix in the circle reduction direcion, {\it not} the time
direction.)  Again, there was nothing incorrect about the definition
used in the old versions of the paper, but it had the consequence that
the two chiral spinors of type IIB were Majorana and anti-Majorana
respectively.  This led to expressions, such as the old versions
of the type IIB Lagrangian (\ref{type2baction}), which although perfectly
correct, were somewhat misleading if not set in proper context.  In this
new version, where we define $\td\theta^2=\Gamma_0\, \theta^2$, the two
chiral spinors of the type IIB theory are both Majorana.  This change of
convention amounts to making the replacement of $\td\theta^2$ in the
old versions of the paper by $-\im\, \td\theta^2$, and of $\bttheta^2$
by $\im\, \bttheta^2$,  in order to get the expressions
we are using in this new version.

   Finally, we have made a simplification that eliminates the need for
certain Dirac-matrix combinations 
\be
S_\mu{}^{\nu_1\cdots \nu_p}\equiv \Gamma_\mu{}^{\nu_1\cdots \nu_p}
  -2p\, \delta_\m^{[\nu_1}\, \Gamma^{\nu_2\cdots \nu_p]}
\ee
that we introduced in the old versions of this paper. It is easily seen
that these are nothing but
\be
S_\mu{}^{\nu_1\cdots \nu_p} = (-1)^p\, \Gamma^{\nu_1\cdots \nu_p}\,
   \Gamma_\mu\,,
\ee
and so we have made these replacements in all occurrences in the old
versions.

\section*{Appendix}

\appendix

\section{Massless/massive IIA T-duality from $D=8$ massive supergravity}

   In this appendix, we discuss the massless/massive IIA T-duality 
at the level of supergravity. 
In order to make explicit the T-duality transformation that
directly maps between the massless and massive type IIA theories, we
shall consider a Scherk-Schwarz dimensional reduction of the massive
type IIA supergravity, in which the 2-form field strength $F_\2$ in $D=10$
is allowed to be non-vanishing in the internal directions of the
2-torus.  Thus in addition to the mass parameter $m_1$ of the original
massive IIA theory, we shall also introduce a second mass parameter
$m_2$ {\it via} the Scherk-Schwarz reduction to $D=8$.  When $m_2=0$,
the $D=8$ theory is the standard two-torus reduction of massive type
IIA, whilst when $m_1=0$, the resuting theory is the Scherk-Schwarz
reduction of massless type IIA or M-theory to $D=8$.

      Our goal is to show that the $m_1=0$ or $m_2=0$ cases are in
fact the same theory, by giving the explicit transformations of the
fields that map one theory to the other. To do so, we shall explicitly
exhibit the symmetry of the resulting eight-dimensional theory under
the $SL(2,\R)$ factor of the $SL(3,\R)\times SL(2,\R)$ Cremmer-Julia
group, under which $(m_1,m_2)$ transform as a doublet.  (We can really
think of $m_1$ and $m_2$ as being fields, rather than just parameters;
see, for example, \cite{llps}.)  In particular, this $SL(2,\R)$
symmetry has a $Z_2$ subgroup that maps from the case where $m_1=0$
with $m_2$ non-vanishing to the case where $m_2=0$ with $m_1$
non-vanishing.  This subgroup, which is the one we found by performing
the double-inversion T-duality transformation (\ref{z2el}) in
section 5, is thus a symmetry that manifestly relates the
Scherk-Schwarz reduction of the massless type IIA theory to the
ordinary reduction of the massive type IIA theory, in eight
dimensions.  This then completes the demonstration of the
massless/massive type IIA T-duality.

       In $D=10$, the massive type IIA theory \cite{romans} is given by 
\bea
{\cal L} &=& R\, {*\oneone} -\ft12 {*d\phi}\wedge d\phi - \ft12
e^{\fft32\phi}\, {*\hat F_\2}\wedge \hat F_\2 - \ft12 e^{-\phi}\,
{*\hat F_\3}\wedge F_\3- \ft12 e^{\fft12\phi}\, {*\hat F_\4}
\wedge F_\4 \nn\\
&&\!\!
-\ft12 d\hat A_\3\wedge d\hat A_\3 \wedge \hat A_\2 -
\ft16 m_1\, d\hat A_\3 \wedge (\hat A_\2)^3
 -\ft1{40} m_1^2\, (\hat A_\2)^5 -\ft12 m_1^2\, e^{\fft52\phi}\, 
{*\oneone}\,,
\label{romans1}
\eea
where the field strengths are given in terms of potentials by
\bea
\hat F_\2 &=& d\hat A_\1 + m_1\, \hat A_\2\ ,\qquad \hat F_\3 =
d\hat A_\2\ ,\nn\\
\hat F_\4 &=& d\hat A_\3 + \hat A_\1\wedge d\hat A_\2
+ \ft12 m_1\, \hat A_\2\wedge \hat A_\2
\,.\label{romfields}
\eea
We now reduce on $T^2$ the standard way, using the conventions and
notation of \cite{cjlp1}, adapted to the case where the first
reduction step is from $D=10$ to $D=9$, and where the generalised 
Scherk-Schwarz reduction is included. To be explicit, we have
\bea
\hat A_\1 &=& A_\1 + A_{\0\a}\, dz^\a - m_2\, z^2\, dz^1\,,\nn\\
\hat A_\2 &=& A_\2 + A_{\1\a}\wedge dz^\a + A_{\0 12}\,
dz^1\wedge dz^2\,,\nn\\
\hat A_\4 &=& A_\3 + A_{\2\a}\wedge dz^\a + A_{\1 12}\wedge
dz^1\wedge dz^2 + \hat A_1\wedge \hat A_\2\,.
\eea
Note that the last term in $\hat A_\4$, which amounts to a field
redefinition, is needed in order that the coordinate $z^2$ does not
appear undifferentiated in the Lagrangian.  The ansatz for the
Einstein-frame metric is $e^{\phi/2}\, d\hat s^2_{10}$ where $d\hat
s^2_{10}$ is the string frame metric given in (\ref{d8red2}).

   We find that the Scherk-Schwarz reduced Lagrangian in $D=8$ is
given by
\bea
{\cal L}_8 &=& R\, {*\oneone} - \ft12{*df}\wedge df 
-\ft12 e^{2f}\, {*d\chi}\wedge d\chi 
-\ft12 {*d\psi}\wedge d\psi 
-\ft12 e^{2\psi}\,{*d\cA^1_{\0 2}}\wedge d\cA^1_{\0 2}\nn\\
&& - \ft12 {*d\sigma}\wedge d\sigma
 -\ft12 e^{-\psi +\sqrt3\sigma}\, {*F_{\1 1}}\wedge F_{\1 1}
-\ft12 e^{\psi +\sqrt3\sigma}\, {*F_{\1 2}}\wedge F_{\1 2}\nn\\
&&-\ft12 e^{-f}\, {*F_\4}\wedge F_\4 -\ft12 e^{-\fft2{\sqrt3}\sigma}\,
{*F_\3}\wedge F_\3   \nn\\
&&
-\ft12 e^{\fft1{\sqrt3} \sigma}\, (e^{-\psi}\,
{*F_{\3 1}}\wedge F_{\3 1} + e^\psi\, {*F_{\3 2}}\wedge F_{\3 2})\nn\\
&&-\ft12 e^{\fft2{\sqrt3}\sigma}\, (e^{-f}\, {*F_\2}\wedge F_\2 +
e^f\, {*F_{\2 12}}\wedge F_{\2 12})\nn\\
&&
-\ft12 e^{-\psi -\fft1{\sqrt3}\sigma}\, (e^{-f}\, {*\cF_\2^2}\wedge 
\cF_\2^2 +e^{f}\, {*F_{\2 1}}\wedge F_{\2 1})\nn\\
&&-\ft12 e^{\psi -\fft1{\sqrt3}\sigma}\,
(e^{-f}\, {*\cF_\2^1}\wedge \cF_\2^1 +
        e^{f}\, {*F_{\2 2}}\wedge F_{\2 2})\nn\\
&&-\ft12 e^{\fft4{\sqrt3}\sigma}\, \Big( m_1^2\, e^{-f} +
  (m_2 +m_1\, \chi)^2\, e^f \Big)\, {*\oneone} + {\cal L}_{FFA}\,,
\label{d8romlag}
\eea
where $\chi=A_{\012}$ and 
\be
{\cal L}_{FFA} = -\ft12\chi\, F_\4\wedge F_\4 +
\Omega_1\wedge dA_\3 + \Omega_2\,,
\ee
with
\bea
\Omega_1 &=&\ep^{\a\b}\,  A_{\1 \a}\, dA_{\2 \b} +\ft12 \ep^{\a\b}\, A_\1\,
A_{\1 \a}\,dA_{\1\b} - A_\2\, dA_{\1 12} \nn\\
&& +\ft12 A_{\1 12}\, A_{\1\a}\, d\cA_\1^\a + A_{\1 1}\, A_{\1 2}\, dA_\1
+ m_1\, A_{\1 1}\, A_{\1 2}\, A_\2 \nn\\
&& - \ft12 m_2 (A_\2)^2 - \ft12 m_2\, A_\2\, A_{\1\a}\, \cA_\1^\a
+\ft14 m_2\, A_{\1 1}\, A_{\1 2}\, \cA_\1^1\, \cA_\1^2\,.
\eea
The expression for $\Omega_2$, which does not involve $A_\3$, is quite 
complicated, and we shall not present it here.   The exterior derivative
of ${\cal L}_{FFA}$ is, however, rather simple, and is given by
\be
d{\cal L}_{FFA} =-\ft12 d(\chi\, F_\4\wedge F_\4) + d\Omega_1\wedge
F_\4 + F_\3\wedge F_{\3 1}\wedge F_{\3 2}\,.
\ee

   In the scalar sector, the $SL(2,\R)$ symmetry is generated by the
axion $\chi\equiv A_{\0 12}$, and the dilaton combination $f$, given
by (\ref{fpsidef}), which arises in the exponential prefactor for the
axion's kinetic term in $D=8$.  To make the $SL(2,\R)$ manifest, it is
useful to perform a number of redefinitions of the
dimensionally-reduced potentials. These are given below, in equation
(\ref{aredefs}).  Expressed in terms of the now-redefined potentials,
we obtain the following expressions for the various field strengths in
$D=8$.  Firstly, we have
\be
F_\3 = dA_\2 -\ft12 A_{\1\a}\, d\cA_\1^\a - \ft12 \cA_\1^\a\, 
dA_{\1\a}\,,\label{f3def}
\ee
which is a singlet under the $SL(2,\R)$.   We find that the remaining
two 3-form field strengths are also $SL(2,\R)$ singlets, and are given by
\bea
F_{\3 1} &=& dA_{\2 1} + \ft12 A_\1\, dA_{\1 1} + \ft12 A_{\1 1}\, dA_\1
+ \ft12 A_{\1 12}\, d\cA_\1^2 + \ft12 \cA_\1^2\, dA_{\1 12} -
A_{\0 1}\, F_\3 \nn\\
&& + A_\2\, (m_1 A_{\1 1} + m_2\, \cA_\1^2) +
 \ft12 A_{\1 1}\, \cA_\1^2\, (m_1\, A_{\1 2} - m_2\, \cA_\1^1)\,,\\
F_{\3 2} &=& dA_{\2 2} + \ft12 A_\1\, dA_{\1 2} + \ft12 A_{\1 2}\, dA_\1
- \ft12 A_{\1 12}\, d\cA_\1^1 - \ft12 \cA_\1^1\, dA_{\1 12} -
A_{\0 2}\, F_\3 \nn\\
&&-\cA^1_{\0 2}\, F_{\3 1}
 + A_\2\, (m_1 A_{\1 2} - m_2\, \cA_\1^1) +
 \ft12 A_{\1 2}\, \cA_\1^1\, (m_1\, A_{\1 1} + m_2\, \cA_\1^2)\,.
\nn
\eea
We also have the pair of singlet 1-forms,
\bea
F_{\1 1} &=& dA_{\0 1} + (m_1\, A_{\1 1} +m_2\, \cA_\1^2) \,,\nn\\
F_{\1 2} &=& dA_{\0 2} -\cA^1_{\0 2}\, F_{\1 1} +
(m_1\, A_{\1 2} - m_2\, \cA_\1^1)\,.
\eea

   Next, we find
\bea
F_\2 &=& dA_\1 +A_{\0\a}\, d\cA_\1^\a
+ m_1\, A_\2 -\ft12  m_1\, A_{\1 \a}\, \cA_\1^\a
+m_2\, \cA_\1^1\, \cA_\1^2\,,\nn\\
F_{\2 12} &=&  dA_{\1 12} + A_{\0 1} \, dA_{\1 2} - 
A_{\0 2}\, dA_{\1 1} + \chi\, F_\2\nn\\
&&+m_2\, A_\2 +\ft12 m_2\, A_\1\, \cA_\1^\a
- m_1\, A_{\1 1}\, A_{\1 2}\,,
\eea
This pair of field strengths forms a doublet under the $SL(2,\R)$ we
are considering.  For the other 2-forms, we find that they form two
pairs, namely
\bea
\cF_\2^2 &=& d\cA_\1^2 \,,\nn\\
 F_{\2 1} &=& dA_{\1 1} -\chi\, d\cA_\1^2
\,,
\eea
and
\bea
\cF_\2^1 &=& d\cA_\1^1 +\cA^1_{\0 2}\, d\cA_\1^2\,,\nn\\
F_{\2 2} &=& dA_{\1 2} + \chi\,
d\cA_\1^1 -\cA^1_{\0 2}\, F_{\2 1}\,.
\eea
Each pair is a doublet under the $SL(2,\R)$.

    Finally, we have the 4-form.  This turns out to be
\bea
F_\4 &=& dA_\3 + A_\1\, dA_\2 + \cA_\1^\a\, dA_{\2\a} -
\ft12 A_\1\, \cA_\1^\a\, dA_{\1\a} \nn\\
&& -\ft12 A_{\1 12}\, (\cA_\1^1\, dA_{\1 2} - \cA_\1^2\, dA_{\1 1})
+\ft12 m_1\, (A_{\2})^2 - \ft12 m_1\, A_{\2}\, A_{\1 \a}\, \cA_\1^\a\nn\\
&& -\ft14 m_1\, A_{\1 1}\, A_{\1 2}\, \cA_\1^1\, \cA_\1^2
+ m_2\, A_\2\, \cA_\1^1\, \cA_\1^2\,.\label{d8f4}
\eea
This, together with its Hodge dual, forms a doublet under the
$SL(2,\R)$.  The symmetry involving $F_\4$ is therefore seen only at
the level of the field equations.

   It is useful at this stage to consider the various doublet
symmetries under the $SL(2,\R)$ generated by $f$ and $\chi$ in more
detail.  First of all, we note from (\ref{d8romlag}) that, as one
would expect, $f$ does not couple to the kinetic terms for any of the
$SL(2,\R)$-singlet fields.  All the doublet pairs of 2-form fields
strengths $(F_+,F_-)$, with potentials $(A_+, A_-)=$ $(A_\1,A_{\1
12})$, $(-\cA_\1^2, A_{\1 1})$ and $(\cA_\1^1, A_{\1 2})$, which we
exhibited above, have kinetic terms which couple to $f$ in the form
$-\ft12 e^{-f}\, F_+^2 -\ft12 e^f\, (F_- + \chi\, F_+)^2$.  Under
$SL(2,\R)$ transformations acting on $\tau=\chi+ \im\, e^{-f}$ {\it
via} the fractional linear mapping $\tau\longrightarrow (a\, \tau +
b)/(c\, \tau + d)$, the potentials $(A_+, A_-)$ transform as
\be
\pmatrix{A_+\cr A_-} \longrightarrow \pmatrix{ d& -c\cr -b & a}\,
\pmatrix{A_+\cr A_-}\,.\label{fpmtrans}
\ee
The two mass parameters $(m_1, m_2)$ transform in the same way as
$(A_+, A_-)$. Finally, to discuss the $SL(2,\R)$ symmetry for $F_\4$,
we note that the pair $F_\4$ and $(e^{-f}\, {*F_\4} + \chi\, F_\4)$
transform like the upper and lower components of a doublet, as in
(\ref{fpmtrans}).  Since this involves the 4-form and its dual, the
$SL(2,\R)$ symmetry is realised here at the level of the equations of
motion rather than the Lagrangian.  The the Bianchi identity and the
equation of motion for the 4-form are given by
\be
dF_\4 = X\,,\qquad d(e^{-f}\, {*F_\4} + \chi\, F_\4) = d\Omega_1\,,
\ee
where $X$ can be easily read off from (\ref{d8f4}).  It is
straightforward to verify that these two equations form a doublet under
the $SL(2,\R)$.

     For completeness, we list the field redefinitions that we made
for the potentials.  Making the following substitutions in the
expressions for the field strengths directly following from the
dimensional reduction, we obtain the field strengths given above:
\bea
A_\3 &\longrightarrow& A_\3 - A_\2\, A_\1, -A_\2\, A_{\0\a}\, \cA_\1^\a +
\ft12 \ep^{\a\b}\, A_{\0\a}\, A_{\1\b}\, \cA_\1^1\, \cA_\1^2\,,\nn\\
A_\2 &\longrightarrow& A_\2 + \ft12 A_{\1\a}\, \cA_\1^\a +
\chi\, \cA_\1^1\, \cA_\1^2\,,\nn\\
A_\1 &\longrightarrow& A_\1 + A_{\0 \a}\, \cA_{\1 \a}\,,\nn\\
A_{\2 1} &\longrightarrow& A_{\2 1} - A_{\0 1}\, A_\2 -
\ft12 A_\1\, A_{\1 1} - \ft12 A_{\1 12}\, \cA_\1^2 \nn\\
&&  +\ft12 A_{\0 1}\, (A_{\1 1}\, \cA_\1^1 - A_{\1 2}\, \cA_\1^2) +
A_{\0 2}\, A_{\1 1}\, \cA_\1^2\,,\nn\\
A_{\2 2} &\longrightarrow& A_{\2 2} - A_{\0 2}\, A_\2 -
\ft12 A_\1\, A_{\1 2} + \ft12 A_{\1 12}\, \cA_\1^1 \nn\\
&&  -\ft12 A_{\0 2}\, (A_{\1 1}\, \cA_\1^1 - A_{\1 2}\, \cA_\1^2) +
A_{\0 1}\, A_{\1 2}\, \cA_\1^1\,,\nn\\
A_{\1 1} &\longrightarrow& A_{\1 1}- \chi\, \cA_\1^2\,,\label{aredefs}\\
A_{\1 2} &\longrightarrow& A_{\1 2} + \chi\, \cA_\1^1\,,\nn\\
A_{\1 12} &\longrightarrow& A_{\1 12} + \ep^{\a\b}\,
A_{\0 \a}\, A_{\1 \b}\,,\nn\\
\cA_\1^1 &\longrightarrow& \cA_\1^1 + \cA^1_{\0 2}\, \cA_\1^2\,.\nn
\eea

     To summarise the appendix, we have constructed a manifestly
$SL(3,\R)\times SL(2,\R)$ invariant massive supergravity in $D=8$,
with two mass parameters $m_1$ and $m_2$ which form a doublet under
the $SL(2,\R)$ factor.  When $m_2=0$ with $m_1$ non-vanishing, the
theory is the standard Kaluza-Klein 2-torus reduction of the massive
type IIA theory in $D=10$.  On the other hand when $m_1=0$ with $m_2$
non-vanishing, the theory comes from the the Scherk-Schwarz reduction
of massless type IIA (or M-theory).  The manifest $SL(2,\R)$ symmetry
implies in particular that these two eight-dimensional cases are
equivalent.  In particular, as we observed in section 5, the discrete
transformation that maps $(m_1, 0)$ to $(0, m_2)$, associated with the
massive/massless type IIA theories, interchanges $R_1 \leftrightarrow
1/R_2$.  This $\Z_2$ is a subgroup of the $SL(2,\R)$; it is quite
different from the $R_1\leftrightarrow 1/R_1$ $\Z_2$ transformation
that relates the type IIA and type IIB theories, which is
intrinsically discrete and is not part of any connected group.

       As was observed in \cite{lav}, the U-duality group can be
decomposed as general coordinate transformations of the internal
spaces of either the type IIA or the type IIB theories, together with
the T-duality transformation that maps between the two theories.  From
the type IIB point of view, the $SL(2,\R)$ symmetry under which $(m_1,
m_2)$ form a doublet is nothing but a residual general coordinate
transformation symmetry of the internal 2-torus. This is because in
the type IIB case, the axion $\chi$ in $D=10$ can be Scherk-Schwarz
reduced on each of the two circles reductions, giving rise to two mass
parameters that naturally form a doublet under the $SL(2,\R)$ residuum
of the 2-torus general coordinate transformations.

\section*{Note added}

    After this work was completed, a paper appeared which also
considers certain aspects of type IIA/IIB T-duality for R-R fields
\cite{new2}.

\section*{Acknowledgements}

     We are grateful to Chris Hull, Igor Lavrinenko and Arkady 
Tseytlin for useful discussions, and to SISSA for hospitality.

\end{document}